\documentclass[a4paper,11pt]{article}
\pdfoutput=1 

\usepackage{jheppub} 

\usepackage[T1]{fontenc} 

\usepackage{mathtools}
\usepackage{tensor}
\usepackage{braket}
\usepackage[compat=1.1.0]{tikz-feynman}
\usepackage{cleveref}
\usepackage{booktabs}
\usepackage{diagbox}
\usepackage{transparent}
\usepackage{color}
\crefname{table}{Table}{Tables}
\crefname{equation}{Eq.}{Eqs.}
\crefname{appendix}{Appendix}{Appendices}
\crefname{section}{Section}{Sections}
\crefname{figure}{Fig.}{Figs.}

\newcommand{\Amp}{\ensuremath{\mathcal{A}}}
\newcommand{\V}{\ensuremath{\mathcal{V}}}
\newcommand{\X}{\ensuremath{X}}

\title{Effective Field Theories as Lagrange Spaces}

\author[a,b]{Nathaniel Craig,}
\author[a,1]{Yu-Tse Lee,\note{Corresponding author.}}
\author[c]{Xiaochuan Lu,}
\author[d]{and Dave Sutherland}

\affiliation[a]{Department of Physics, University of California, Santa Barbara, CA 93106, USA}
\affiliation[b]{Kavli Institute for Theoretical Physics, Santa Barbara, CA 93106, USA}
\affiliation[c]{Department of Physics, University of California, San Diego, La Jolla, CA 92093, USA}
\affiliation[d]{School of Physics and Astronomy, University of Glasgow, Glasgow G12 8QQ, United Kingdom}

\emailAdd{ncraig@ucsb.edu}
\emailAdd{yutselee@ucsb.edu}
\emailAdd{xil224@ucsd.edu}
\emailAdd{david.w.sutherland@glasgow.ac.uk}

\abstract{We present a formulation of scalar effective field theories in terms of the geometry of Lagrange spaces. The horizontal geometry of the Lagrange space generalizes the Riemannian geometry on the scalar field manifold, inducing a broad class of affine connections that can be used to covariantly express and simplify tree-level scattering amplitudes. Meanwhile, the vertical geometry of the Lagrange space characterizes the physical validity of the effective field theory, as a torsion component comprises strictly higher-point Wilson coefficients. Imposing analyticity, unitarity, and symmetry on the theory then constrains the signs and sizes of derivatives of the torsion component, implying that physical theories correspond to a special class of vertical geometry.}

\begin{document}
\maketitle
\flushbottom

\section{Introduction}
\label{sec:intro}

The Lagrangian formulation of relativistic effective field theories (EFTs) suffers from a redundancy of description associated with the field redefinition invariance of scattering amplitudes~\cite{Kamefuchi:1961sb, Chisholm:1961tha, Coleman:1969sm,Arzt:1993gz}. The insensitivity of EFT observables to field parameterization is reminiscent of coordinate invariance in general relativity, suggesting that geometry should be a powerful tool in both settings. Indeed, it has long been appreciated that scalar field theories admit a geometric description in which the fields play the role of coordinates on the field space manifold \cite{Meetz:1969as, Honerkamp:1971sh, Honerkamp:1971xtx, Ecker:1971xko, Alvarez-Gaume:1981exa, Alvarez-Gaume:1981exv, Boulware:1981ns, Howe:1986vm}, and scattering amplitudes can be expressed in terms of corresponding geometric quantities that reflect the underlying parameterization invariance \cite{Dixon:1989fj, Alonso:2015fsp}. Not only does geometry naturally organize EFT observables, it also reveals deeper properties (such as the linearly realized symmetries) that can otherwise be obscured at the level of the Lagrangian \cite{Alonso:2015fsp, Cohen:2020xca}.

The geometric approach to EFT has proven particularly fruitful in characterizing deviations from the Standard Model~\cite{Alonso:2016oah, Nagai:2019tgi, Helset:2020yio, Cohen:2021ucp, Alonso:2021rac}. Recent work has also explored the effects of loop corrections and renormalization on the Riemannian field space geometry~\cite{Alonso:2022ffe, Helset:2022tlf, Helset:2022pde}, as well as extensions to higher-spin theories~\cite{Finn:2019aip, Finn:2020nvn}. However, Riemannian field space geometry is not without its limitations. The geometry of the scalar manifold is wholly determined by the two-derivative part of the action, rendering it insensitive to the properties of higher-derivative terms that encode essential physical information about analyticity, unitarity, and locality of the EFT. It also fails to reflect the full invariance of scattering amplitudes by encoding only invariance under non-derivative field redefinitions. Significant progress has recently been made on the development of more general structures accommodating derivative field redefinitions~\cite{Cohen:2022uuw, Cheung:2022vnd}, but they involve significantly abstracted infinite-dimensional manifolds. If geometry is to fully capture the structure of effective field theories, a new approach is required.

\begin{figure}[tbp]
\centering
\def\svgwidth{\columnwidth}
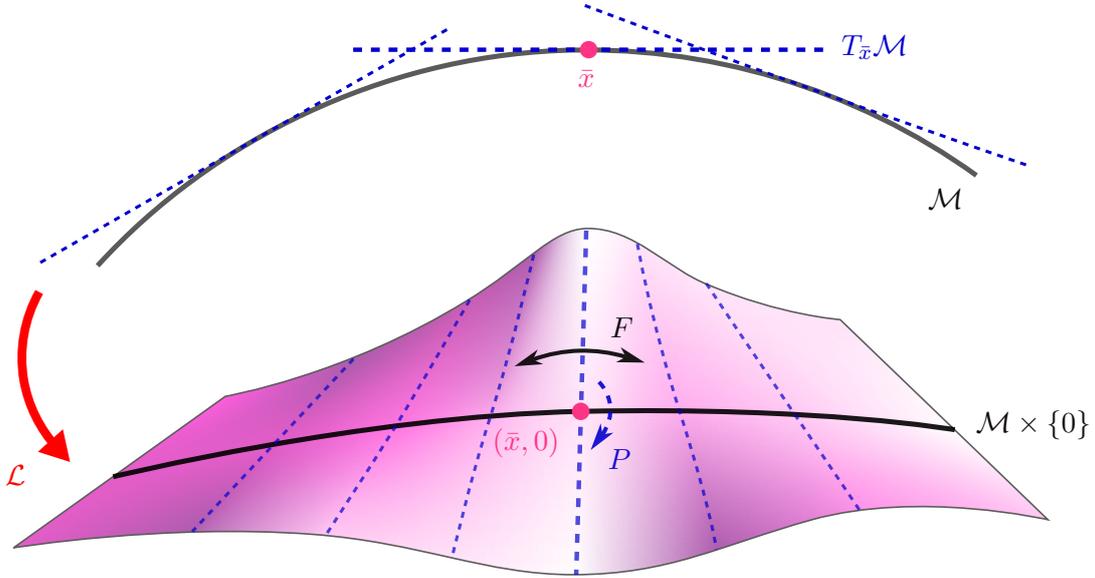
\caption{\label{fig:lagrangefig}An illustration of the field theory Lagrange space (not drawn to mathematical accuracy). The tangent bundle $T\mathcal{M}$ of the scalar field manifold $\mathcal{M}$ is built from the tangent space (dashed line) at each point on the manifold (solid line). Endowing $T\mathcal{M}$ with a Lagrangian $\mathcal{L}$ makes it a Lagrange space, whose geometry can be decomposed as horizontal (``parallel to the solid line'') or vertical (``parallel to the dashed lines''). At a point $(\bar{x}, 0)$ which we call the vacuum, the horizontal geometry characterized by the h-Christoffel symbols $F$ describes scattering amplitudes in a fashion similar to --- but more general than --- the Riemannian framework, whereas the vertical geometry characterized by the torsion $P$ describes strictly higher-order physics of the field theory.}
\end{figure}

In this work we take a first natural step towards a more comprehensive EFT geometry by incorporating derivative coordinates and enlarging the geometric setting to include the tangent bundle of the scalar field manifold, a substantial generalization that nevertheless maintains an immediate relationship with the conventional Riemannian framework. We find that a generic scalar Lagrangian with symmetric Wilson coefficients can be naturally and wholly embedded in the tangent bundle, giving rise to what is known as a {\it Lagrange space} \cite{Kern1974}. Lagrange spaces have been studied in the context of differential equations and the calculus of variations~\cite{Miron:1994nvt, Antonelli:1996dq}. Notably, their geometry can be canonically decomposed as horizontal or vertical. The horizontal geometry induces a broad class of affine connections on the original scalar field manifold. This includes not only the familiar Riemannian construction, but also alternative connections that intrinsically contain additional information on scattering at higher orders in momentum. The vertical geometry can be identified solely with higher-derivative operators in the Lagrangian. Constraints based on physical principles, such as positivity~\cite{Pham:1985cr, Ananthanarayan:1994hf, Adams:2006sv} and unitarity bounds~\cite{Cornwall:1974km,Lee:1977eg, Lee:1977yc, Dicus:1973gbw, Chanowitz:1985hj}, then establish a correspondence between the validity of effective field theories and classes of Lagrange spaces with a special vertical geometry. Thus, unlike the existing Riemannian framework, Lagrange spaces are able to characterize higher-derivative physics as geometry.  This is summarized in the schematic illustration in \cref{fig:lagrangefig}. While Lagrange spaces meaningfully incorporate higher-derivative physics as geometry, it bears emphasizing that they do not capture the invariance of scattering amplitudes under derivative field redefinitions. Nonetheless, the inclusion of derivative coordinates already brings in new physics into the EFT geometry.

The paper is organized as follows: In \cref{sec:lagrange} we introduce the geometry of Lagrange spaces and construct one from the Lagrangian of a generic scalar EFT. In \cref{sec:horgeom} we express scattering amplitudes covariantly using the horizontal geometry of the Lagrange space, revealing that some higher-order derivative physics is more naturally described by connections other than the Riemannian one. We turn to the vertical geometry of the Lagrange space in \cref{sec:vergeom}, using the implications of unitarity and analyticity to establish constraints on a vertical torsion component based on generic physical principles. Lastly, we summarize our results and suggest directions for future investigation in \cref{sec:conc}.

\section{A Generalized Geometric Framework for Effective Field Theories}
\label{sec:lagrange}

It is well understood that tree-level scattering amplitudes in scalar quantum field theories can be written as covariant tensors on a Riemannian scalar field manifold. As a review, let us adopt the mostly minus spacetime metric and consider a theory of scalar fields $\phi^\alpha$ labeled by flavor indices $\alpha$. The scalar field manifold $\mathcal{M}$ is one that is charted by $\phi$, with dimension equal to the number of flavors, and coordinate transformations on $\mathcal{M}$ are non-derivative field redefinitions $\phi(\tilde{\phi})$.\footnote{For notational brevity, flavor indices $\{\alpha_i\}$ on tensors may be dropped after they are introduced, as we have done here.} A two-derivative Lagrangian can be written generically as\footnote{We have adopted an unconventional sign for the potential term $V(\phi)$, as well as a different normalization of the function $g_{\alpha\beta}(\phi)$ compared to e.g. Refs.~\cite{Alonso:2015fsp, Alonso:2016oah, Cohen:2021ucp}.}
\begin{equation}
    \mathcal{L}(\phi, \partial_\mu \phi) = V(\phi) + g_{\alpha \beta} (\phi) (\partial_\mu \phi^\alpha) (\partial^\mu \phi^\beta) + \mathcal{O}(\partial^4) \,.
\label{eqn:Lagp2}
\end{equation}
Since the Lagrangian is invariant under field redefinitions, the following transformation rules hold for terms of each order in $\left(\partial_\mu \phi\right)$:
\begin{subequations}
\begin{align}
    \tilde{V}(\tilde{\phi}) &= V \big( \phi(\tilde{\phi}) \big) \,, \\[5pt]
    \tilde{g}_{\alpha \beta} (\tilde{\phi}) &= g_{\gamma \delta} \big( \phi(\tilde{\phi}) \big)\, \frac{\partial \phi^\gamma(\tilde{\phi})}{\partial \tilde{\phi}^\alpha} \frac{\partial \phi^\delta(\tilde{\phi})}{\partial \tilde{\phi}^\beta} \,\,.
\end{align}
\end{subequations}
Evidently, the potential $V(\phi)$ is a scalar function and $g_{\alpha\beta}(\phi)$ is a symmetric 2-tensor on $\mathcal{M}$. Moreover, $g_{\alpha\beta}(\phi)$ must be positive definite as a matrix in a unitary theory and hence can be viewed as a Riemannian metric on $\mathcal{M}$. There is then a metric-compatible Levi-Civita connection with the usual Christoffel symbols
\begin{equation}
    \Gamma{^\alpha}_{\beta\gamma} = \frac12\, g^{\alpha\lambda}
    \left( g_{\lambda\beta, \gamma} + g_{\lambda\gamma, \beta} - g_{\beta\gamma, \lambda} \right) \,,
\label{eqn:LeviCivitaConnection}
\end{equation}
\noindent and Riemann curvature tensor
\begin{equation}
    R{^\alpha}_{\beta\gamma\delta} = \Gamma{^\alpha}_{\beta\delta,\gamma} + \Gamma{^\alpha}_{\lambda\gamma} \Gamma{^\lambda}_{\beta\delta} - \left( \gamma \leftrightarrow \delta \right) \,.
\label{eqn:RiemannTensor}
\end{equation}
We will write partial derivatives as commas like above, and Levi-Civita covariant derivatives as semi-colons like below.

Since these geometric structures contain information from the Lagrangian, it is not a far stretch to postulate that their derived quantities contain physical meaning. Indeed, it turns out that tree-level scattering amplitudes can be expressed using such quantities at each order in kinematics~\cite{Cohen:2021ucp}: 
\begin{align}
    \left ( \prod_{i=1}^n \sqrt{2 \bar{g}_{\alpha_i \alpha_i}} \right ) \Amp_n
    &= \bar{V}_{;(\alpha_1 \ldots \alpha_n)} - 2\,\frac{n-3}{n-1}\, \sum_{i<j} s_{ij} \left [ \bar{R}_{\alpha_i (\alpha_1 \alpha_2 | \alpha_j ; | \alpha_3 \ldots \alpha_n)} + \mathcal{O}(\bar{R}^2) \right ]
    \notag\\[5pt]
    &\quad
    + \text{ factorizable pieces} \,,
\label{eqn:AnptLC}
\end{align}
where the factorizable pieces are rational terms with one or more poles in $s_{ij}$, that can be unambiguously constructed from expressions for lower-point amplitudes as discussed in \cite{Cohen:2021ucp}. Let us unpack this expression:
\begin{itemize}
\item Here, $\Amp_n$ is the scattering amplitude of $n$ particles each of flavor $\alpha_i$ and ingoing momentum $p_i$; the Mandelstam variables are $s_{ij} = (p_i + p_j)^2$.
\item Barred quantities are to be evaluated at the physical vacuum $\bar{\phi} \in \mathcal{M}$ that minimizes the potential $V(\phi)$. Parentheses are used to denote (normalized) symmetrizations of the flavor indices, and short vertical bars are used to demarcate segments of indices that are not involved in the symmetrization.
\item Coordinates that diagonalize $\bar{V}_{,\alpha\beta}$ and $\bar{g}_{\alpha\beta}$ with ratios $\bar{V}_{,\alpha\beta} = -2\, \bar{g}_{\alpha\beta}\, m^2_\alpha$ have been adopted without loss of generality, so that wavefunction normalization factors can be expressed as diagonal entries, even though these are really eigenvalues of the matrix $\bar{g}_{\alpha\beta}$.
\item Some terms have been made implicit for brevity, because they are either of higher orders in curvature, or can be decomposed into lower-point amplitudes glued together by propagator factors.
\end{itemize}
As a concrete example of \cref{eqn:AnptLC}, take the four-point amplitude $\Amp_4$:
\begin{align}
\left( \prod_{i=1}^4 \sqrt{2 \bar{g}_{\alpha_i\alpha_i}} \right) \Amp_4
&= \bar{V}_{; (\alpha_1 \alpha_2 \alpha_3 \alpha_4)} - \frac23\, \sum_{i<j} s_{ij} \bar{R}_{\alpha_i (\alpha_k\alpha_l) \alpha_j}
\notag\\[5pt]
&\quad
- \bar{V}_{; (\alpha_1\alpha_2\beta)}\, \frac{\bar{g}^{\beta\gamma}}{2(s_{12} - m_\beta^2)}\,  \bar{V}_{; (\alpha_3\alpha_4\gamma)}
- \bar{V}_{; (\alpha_1\alpha_3\beta)}\, \frac{\bar{g}^{\beta\gamma}}{2(s_{13} - m_\beta^2)}\,  \bar{V}_{; (\alpha_2\alpha_4\gamma)}
\notag\\[5pt]
&\quad
- \bar{V}_{; (\alpha_1\alpha_4\beta)}\, \frac{\bar{g}^{\beta\gamma}}{2(s_{14} - m_\beta^2)}\,  \bar{V}_{; (\alpha_2\alpha_3\gamma)} \,.
\label{eqn:A4ptLC}
\end{align}
Here the factorizable pieces have been written out for explicitness, and the dummy flavor indices $\beta$ and $\gamma$ are to be summed over all flavors of the scalar fields. At four-point, there are no $\mathcal{O}(\bar{R}^2)$ pieces; they only begin to show up from six-point onwards. Note that the propagators are in general the inverse matrices of what is in the parentheses below, arising from the kinetic term in the Lagrangian:
\begin{equation}
\mathcal{L} \supset \frac12\, \phi^\beta \left( -2\, \bar{g}_{\beta\gamma}\, \partial^2 + \bar{V}_{,\beta\gamma} \right) \phi^\gamma \,.
\end{equation}
Since we are adopting coordinates that make this matrix diagonal, we will simply write the propagators as diagonal entries like in the above.

The success of this approach suggests the question: are there alternative ways of encoding the physics of effective field theories as geometry? It is worth highlighting from \cref{eqn:AnptLC} that the leading-order term in momentum, i.e. the order-$s$ term, is fully captured by the curvature tensor (and its covariant derivatives) alone, determined solely by the Riemannian geometry following from $g_{\alpha \beta} (\phi)$. However, the remaining information about the Lagrangian has been relegated to other tensors on $\mathcal{M}$ like the potential $V(\phi)$. Can we meaningfully encode the Lagrangian in just a single structure on some manifold? An answer arises upon a quick detour into the geometric formulation of classical mechanics --- Lagrange spaces.

\subsection{What are Lagrange Spaces?}

Lagrange spaces are the natural geometric setting for the Lagrangian formulation of classical mechanics. The natural variables in classical mechanics comprise the positions and velocities of a system, which we denote as $x^\alpha$ and $y^\beta = \partial_t\, x^\beta$ respectively. Consider the space of all positions the system can take. This is a manifold charted by $x$, which we call $\mathcal{M}$. The possible velocities at each point $x \in \mathcal{M}$ by definition form the tangent space $T_x \mathcal{M}$. The collection of tangent spaces, known as the tangent bundle $T\mathcal{M}$, is then a manifold charted by $(x, y)$, with twice the dimension of $\mathcal{M}$. The slice $\mathcal{M} \times \{ 0 \} = \{ (x, 0) \, | \, x \in \mathcal{M} \} \cong \mathcal{M}$ is termed the null section.

A time-independent classical mechanics Lagrangian $\mathcal{L}(x, y)$ can be interpreted as a smooth real-valued function on $T\mathcal{M}$. Let us additionally assume that $\mathcal{L}$ is regular --- the fundamental tensor
\begin{equation}
\mathfrak{g}_{\alpha \beta}(x, y) \coloneqq \frac12\, \frac{\partial^2}{\partial y^\alpha \partial y^\beta}\, \mathcal{L}(x, y) \,,
\label{eqn:gFundamental}
\end{equation}
has full rank on $T\mathcal{M}$.\footnote{In truth, it is sufficient for our purposes that the fundamental tensor has full rank at (and hence near, by lower semi-continuity) a point we will also call the vacuum. Our requirements differ from mathematics literature, which does not demand that the Lagrangian is smooth and regular on the null section.} Then $\mathcal{M}$ equipped with $\mathcal{L}$ is known as a Lagrange space~\cite{Miron:1994nvt}. The dynamics of the system are embedded into the Lagrange space via what is known as a semispray, whose coefficients are
\begin{equation}
G^\alpha \coloneqq \frac14\, \mathfrak{g}^{\alpha\gamma} \left( \frac{\partial^2 \mathcal{L}}{\partial y^\gamma \partial x^\delta}\, y^\delta - \frac{\partial \mathcal{L}}{\partial x^\gamma} \right) \,.
\label{eqn:Semispray}
\end{equation}
They arise from the Euler-Lagrange equations
\begin{equation}
\partial_t^2\, x^\alpha + 2\, G^\alpha = 0 \,.
\end{equation}
Note the use of the fundamental tensor $\mathfrak{g}_{\alpha \beta}$ and its inverse to lower and raise indices, which requires a regular Lagrangian.

While understanding classical dynamics on Lagrange spaces is an interesting endeavor on its own, there are other remarkable structures that follow from the semispray $G^\alpha$. Consider a coordinate transformation $x(\tilde{x})$ on $\mathcal{M}$. The Lagrangian, a function on $T\mathcal{M}$, will depend on the new coordinates through
\begin{equation}
\mathcal{L}\, \big(\, x(\tilde{x}) \;,\; y(\tilde{x}, \tilde{y}) \,\big) \,.
\end{equation}
The chain rule yields the following induced transformation rules of other objects on $T\mathcal{M}$:
\begin{subequations}\label{eqn:InducedTrans}
\begin{alignat}{3}
x^\alpha &= x^\alpha(\tilde{x}^\beta) \,,\qquad
dx^\alpha &&= \frac{\partial x^\alpha}{\partial \tilde{x}^\beta} d\tilde{x}^\beta \,,\qquad
&&\frac{\partial}{\partial \tilde{x}^\alpha} = \frac{\partial x^\beta}{\partial \tilde{x}^\alpha} \frac{\partial}{\partial x^\beta} + \frac{\partial y^\beta}{\partial \tilde{x}^\alpha} \frac{\partial}{\partial y^\beta} \,, \\[5pt]
y^\alpha &= \frac{\partial x^\alpha}{\partial \tilde{x}^\beta} \tilde{y}^\beta \,,\qquad
dy^\alpha &&= \frac{\partial x^\alpha}{\partial \tilde{x}^\beta} d\tilde{y}^\beta + \frac{\partial y^\alpha}{\partial \tilde{x}^\beta} d\tilde{x}^\beta \,,\qquad
&&\frac{\partial}{\partial \tilde{y}^\alpha} = \frac{\partial x^\beta}{\partial \tilde{x}^\alpha} \frac{\partial}{\partial y^\beta} \,.
\end{alignat}
\end{subequations}
Let us define a distinguished tensor (or d-tensor) on $T\mathcal{M}$ to be an object $T\indices{^{\alpha\ldots}_{\beta\ldots}} (x,y)$ whose indices run from $1$ to $\dim \mathcal{M}$ (not $\dim T\mathcal{M}$), and whose transformation rule under an induced transformation $x(\tilde{x})$ is
\begin{equation}
T\indices{^\alpha^\ldots_\beta_\ldots} = \left ( \frac{\partial x^\alpha}{\partial \tilde{x}^\gamma} \ldots \right ) \tilde{T}\indices{^\gamma^\ldots_\delta_\ldots} \left( \frac{\partial \tilde{x}^\delta}{\partial x^\beta} \ldots \right) \,.
\label{eqn:dtensor}
\end{equation}
In other words, d-tensors live on $T\mathcal{M}$ but transform like tensors on $\mathcal{M}$. Of the objects in \cref{eqn:InducedTrans}, $y$, $dx$ and $\partial / \partial y$ are d-tensors while $dy$ and $\partial / \partial x$ are not. We can fix the latter by introducing the combinations
\begin{subequations}
\begin{align}
\delta y^\alpha &\coloneqq dy^\alpha + N\indices{^\alpha_\beta} dx^\beta \,, \\[5pt]
\frac{\delta}{\delta x^\beta} &\coloneqq \frac{\partial}{\partial x^\beta} - N\indices{^\alpha_\beta} \frac{\partial}{\partial y^\alpha} \,,
\end{align}
\end{subequations}
with
\begin{equation}
N\indices{^\alpha_\beta} = \frac{\partial G^\alpha}{\partial y^\beta} \,,
\label{eqn:Ndef}
\end{equation}
from the semispray. The coefficients $N\indices{^\alpha_\beta}$ comprise what is known as a non-linear connection on $T\mathcal{M}$, while the sets $\{ \delta / \delta x, \partial / \partial y \}$ and $\{ dx, \delta y \}$ are known as the Berwald bases for the tangent and cotangent bundles of $T\mathcal{M}$, decomposing them into horizontal (those with $x$'s) and vertical ($y$'s) sub-bundles.

\begin{table}[tbp]
\renewcommand{\arraystretch}{1.5}
\setlength{\arrayrulewidth}{.2mm}
\setlength{\tabcolsep}{1em}
\centering
\begin{tabular}{c|cc}
	\hline\hline
    \diagbox[width=5.5cm,height=2.5\line]{$\qquad F\indices{^\alpha_\beta_\gamma}$}{$C\indices{^\alpha_\beta_\gamma}$} & $\dfrac12\, \mathfrak{g}^{\alpha \delta} \left(\dfrac{\partial \mathfrak{g}_{\delta\gamma}}{\partial y^\beta} + \dfrac{\partial \mathfrak{g}_{\delta\beta}}{\partial y^\gamma} - \dfrac{\partial \mathfrak{g}_{\beta\gamma}}{\partial y^\delta}\right)$ & $0$ \\
    \hline
    \rule{0pt}{5ex} $\dfrac12\, \mathfrak{g}^{\alpha \delta} \left( \dfrac{\delta \mathfrak{g}_{\delta\gamma}}{\delta x^\beta} + \dfrac{\delta \mathfrak{g}_{\delta\beta}}{\delta x^\gamma} - \dfrac{\delta \mathfrak{g}_{\beta\gamma}}{\delta x^\delta} \right)$ & Cartan & Chern-Rund \\[10pt]
    \hline
    \rule{0pt}{5ex} $\dfrac{\partial N\indices{^\alpha_\gamma}}{\partial y^\beta}$ & Hashiguchi & Berwald \\[10pt]
    \hline\hline
\end{tabular}
\caption{Common choices of h- and v-Christoffel symbols $F\indices{^\alpha_\beta_\gamma}$ and $C\indices{^\alpha_\beta_\gamma}$ for the $N$-linear connection $\nabla$ defined in \cref{eqn:Nlinear}.}
\label{tab:Connections}
\end{table}

Still, there is more to be asked for --- can we, for instance, transport a horizontal tangent vector from one point to another such that it remains horizontal? This is possible if we define an additional affine connection $\nabla$ on $T\mathcal{M}$ that satisfies
\begin{subequations}\label{eqn:Nlinear}
\begin{alignat}{2}
\nabla_{\delta / \delta x^\gamma} \frac{\delta}{\delta x^\beta} &= F\indices{^\alpha_\beta_\gamma} \frac{\delta}{\delta x^\alpha} \,,\qquad
\nabla_{\delta / \delta x^\gamma} \frac{\partial}{\partial y^\beta} &&= F\indices{^\alpha_\beta_\gamma} \frac{\partial}{\partial y^\alpha} \,, \\[5pt]
\nabla_{\partial / \partial y^\gamma} \frac{\delta}{\delta x^\beta} &= C\indices{^\alpha_\beta_\gamma} \frac{\delta}{\delta x^\alpha} \,,\qquad 
\nabla_{\partial / \partial y^\gamma} \frac{\partial}{\partial y^\beta} &&= C\indices{^\alpha_\beta_\gamma} \frac{\partial}{\partial y^\alpha} \,.
\end{alignat}
\end{subequations}
This so-called $N$-linear connection $\nabla$ is specified by the horizontal (or h-) and vertical (or v-) Christoffel symbols $F\indices{^\alpha_\beta_\gamma}$ and $C\indices{^\alpha_\beta_\gamma}$, which are required to transform like Christoffel symbols and $(1,2)$-tensors on $\mathcal{M}$ respectively. A few common choices in the literature are listed in \cref{tab:Connections}.

The covariant derivative of the $N$-linear connection $\nabla$ on a general d-tensor $T$ decomposes into h- and v-covariant ones, which we denote using a short slash and a tall vertical bar respectively:\footnote{Again we deviate from mathematics literature, this time in notation. This is because we have reserved the short vertical bars for denoting the exclusion of flavor index segments from symmetrization, as was done in e.g. \cref{eqn:AnptLC}.}
\begin{subequations}
\begin{align}
(\nabla T)\indices{^{\alpha \ldots}_{\beta \ldots}} &= T\indices{^{\alpha\ldots}_{\beta\ldots / \mu}}\, dx^\mu + T\indices{^{\alpha \ldots}_{\beta\ldots}} |_\mu\, \delta y^\mu \,,
\end{align}
with
\begin{align}
T\indices{^{\alpha\ldots}_{\beta\ldots / \mu}} &= \frac{\delta}{\delta x^\mu}\, T\indices{^{\alpha \ldots}_{\beta \ldots}} + \left( F\indices{^\alpha_{\gamma\mu}} T\indices{^{\gamma \ldots}_{\beta \ldots}} + \ldots \right) - \left( F\indices{^\gamma_{\beta\mu}} T\indices{^{\alpha \ldots}_{\gamma \ldots}} + \ldots \right) \,, \label{eqn:hD} \\[8pt]
T\indices{^{\alpha \ldots}_{\beta\ldots}} |_\mu &= \frac{\partial}{\partial y^\mu}\, T\indices{^{\alpha \ldots}_{\beta \ldots}} + \left( C\indices{^\alpha_{\gamma\mu}} T\indices{^{\gamma \ldots}_{\beta \ldots}} + \ldots \right) - \left( C\indices{^\gamma_{\beta\mu}} T\indices{^{\alpha \ldots}_{\gamma \ldots}} + \ldots \right) \,. \label{eqn:vD}
\end{align}
\end{subequations}

With an affine connection comes two fundamental invariants --- curvature and torsion. We leave the enumeration of their components to \cref{app:torcur}, but highlight two that will be of vital importance, the hh-curvature $\mathcal{R}$ and the (v)hv-torsion $P$:\footnote{The prefixes mean that $\mathcal{R}$ is the component of the curvature with $\gamma$ and $\delta$ horizontal, and $P$ is the component of the torsion with $\alpha$, $\beta$ and $\gamma$ vertical, horizontal, and vertical respectively.}
\begin{subequations}
\begin{align}
\mathcal{R}\indices{^\alpha_{\beta\gamma\delta}} &\coloneqq dx^\alpha \left ( \left ( \left [\nabla_{\delta / \delta x^\gamma} , \nabla_{\delta / \delta x^\delta} \right ] - \nabla_{[\delta / \delta x^\gamma, \delta / \delta x^\delta]} \right ) \frac{\delta}{\delta x^\beta} \right ) \notag\\[5pt]
&= \frac{\delta F\indices{^\alpha_\beta_\delta}}{\delta x^\gamma} + F\indices{^\alpha_\epsilon_\gamma} F\indices{^\epsilon_\beta_\delta} + C\indices{^\alpha_\beta_\epsilon} \frac{\delta N^\epsilon_{\;\delta}}{\delta x^\gamma} - (\gamma \leftrightarrow \delta) \,, \\[10pt]
P\indices{^\alpha_\beta_\gamma} &\coloneqq \delta y^\alpha \left ( \nabla_{\delta / \delta x^\beta} \frac{\partial}{\partial y^\gamma} - \nabla_{\partial / \partial y^\gamma} \frac{\delta}{\delta x^\beta} - \left [\frac{\delta}{\delta x^\beta}, \frac{\partial}{\partial y^\gamma} \right ] \right ) \notag\\[5pt]
&= F\indices{^\alpha_\gamma_\beta} - \frac{\partial N\indices{^\alpha_\beta}}{\partial y^\gamma} \,.
\end{align}
\end{subequations}

\subsection{The Lagrange Space of an Effective Field Theory}

Let us return to quantum field theory, with attention to an otherwise generic scalar Lagrangian which contains only first derivatives in fields and symmetric Wilson coefficients: 
\begin{align}
\mathcal{L}(\phi, \partial_\mu \phi) &= V(\phi) + g_{\alpha \beta}(\phi)\, (\partial_\mu \phi^\alpha) (\partial^\mu \phi^\beta) \notag\\[5pt]
&\quad + \sum_{k \geq 2} c_{\gamma_1 \ldots \gamma_{2k}}(\phi)\, (\partial_{\mu_1} \phi^{\gamma_1}) (\partial^{\mu_1} \phi^{\gamma_2}) \cdots (\partial_{\mu_k} \phi^{\gamma_{2k-1}}) (\partial^{\mu_k} \phi^{\gamma_{2k}}) \,,
\label{eqn:Lagwithc}
\end{align}
where
\begin{equation}
c_{\gamma_1 \ldots \gamma_{2k}} = c_{(\gamma_1 \ldots \gamma_{2k})}
\quad\text{for all}\quad
k \geq 2 \,.
\end{equation}
Note the restrictions necessary for the map to Lagrange space: with no second or higher derivatives, the Lagrangian is a function of $x$ and $y$. The $y$ coordinate cannot distinguish spacetime derivatives $\partial_\mu \phi$ with different indices $\mu$, meaning the Lagrange space cannot distinguish between $\partial_\mu \phi^{\gamma_1}\partial^\mu \phi^{\gamma_2} \partial_\nu \phi^{\gamma_3}\partial^\nu \phi^{\gamma_4}$ and $\partial_\mu \phi^{\gamma_1}\partial^\mu \phi^{\gamma_3} \partial_\nu \phi^{\gamma_2}\partial^\nu \phi^{\gamma_4}$. Thus, we symmetrize the flavor indices of four- and higher-derivative operators.\footnote{We note here another geometric approach to classical mechanics, the Eisenhart lift, which has been applied to scalar field theories in \cite{Finn:2018cfs}. In \cite{Finn:2018cfs} the generalization from time to spacetime derivatives is handled by the inclusion of auxiliary vector fields.}

The field derivatives transform as $\partial_\mu \phi^\alpha = (\partial \phi^\alpha / \partial \tilde{\phi}^\beta) (\partial_\mu \tilde{\phi}^\beta)$ under a non-derivative field redefinition $\phi(\tilde{\phi})$, and they always come in pairs with spacetime indices contracted. A comparison with \cref{eqn:InducedTrans} then reveals that we can make the identification
\begin{equation}
\phi^\alpha \rightarrow x^\alpha \,,\qquad 
(\partial_\mu \phi^\alpha) (\partial^\mu \phi^\beta) \rightarrow y^\alpha y^\beta \,,
\end{equation}
and encode the Lagrangian with no loss of information as a function
\begin{equation}
\mathcal{L}(x, y) = V(x) + g_{\alpha \beta}(x)\, y^\alpha y^\beta + \sum_{k \geq 2} c_{\gamma_1 \ldots \gamma_{2k}} (x)\, y^{\gamma_1} \cdots y^{\gamma_{2k}} \,,
\label{eqn:Lagy}
\end{equation}
on the tangent bundle $T\mathcal{M}$ of the scalar field manifold. Henceforth, $\phi$ and $x$ will be used interchangeably, the latter not to be confused with spacetime coordinates.

For this Lagrangian, the fundamental tensor defined in \cref{eqn:gFundamental} is given by
\begin{equation}
\mathfrak{g}_{\alpha\beta}(x, y) = g_{\alpha\beta}(x) + \sum_{k \geq 2} k \left(2k-1\right) c_{\alpha\beta \gamma_1 \ldots \gamma_{2k-2}}(x)\, y^{\gamma_1} \cdots y^{\gamma_{2k-2}} \,.
\label{eqn:gFundamentaly}
\end{equation}
One can similarly work out the semispray $G^\alpha$ defined in \cref{eqn:Semispray} and the non-linear connection $N\indices{^\alpha_\beta} = \partial G^\alpha / \partial y^\beta$. In general, we see that these geometric quantities on the Lagrange space $(\mathcal{M}, \mathcal{L})$ contain all higher-order derivative interactions $c_{\gamma_1 \ldots \gamma_{2k}}(x)$. This is in contrast with geometric quantities on the Riemannian space $(\mathcal{M}, g(x))$, such as the Levi-Civita Christoffel symbols $\Gamma\indices{^\alpha_\beta_\gamma}(x)$ or the curvature $R_{\alpha\beta\gamma\delta}(x)$, which involve only the two-derivative term $g_{\alpha\beta}(x)$.

Now let us examine the null-section ($y=0$) properties of these geometric quantities on the Lagrange space $(\mathcal{M}, \mathcal{L})$. Since the Lagrangian specified in \cref{eqn:Lagy} is an even function in $y$, both the fundamental tensor $\mathfrak{g}_{\alpha\beta}(x,y)$ and the semispray $G^\alpha(x,y)$ will also be even, following \cref{eqn:gFundamental,eqn:Semispray}. As a consequence, the non-linear connection $N{^\alpha}_\beta(x,y)$ defined in \cref{eqn:Ndef} is odd and vanishes on the null section: $N{^\alpha}_\beta(x,0)=0$. This has an important consequence --- the horizontal part of an $N$-linear connection $\nabla$ on $T\mathcal{M}$, when restricted to the null section, reduces to an affine connection on $\mathcal{M}$. To see why, consider a general d-tensor $T{^{\alpha\ldots}}_{\beta\ldots} (x,y)$. Its horizontal covariant derivative $T{^{\alpha\ldots}}_{\beta\ldots / \mu} (x,y)$ follows from \cref{eqn:hD}. Focusing on its null-section value, we find
\begin{align}
T{^{\alpha\ldots}}_{\beta\ldots / \mu} (x,0) &= \left( \frac{\partial}{\partial x^\mu}\, T{^{\alpha\ldots}}_{\beta\ldots} - N{^\lambda}_\mu \frac{\partial}{\partial y^\lambda}\, T{^{\alpha\ldots}}_{\beta\ldots} \right) \biggr|_{y=0}
+ \Big[ F{^\alpha}_{\gamma\mu}(x,0)\, T{^{\gamma\ldots}}_{\beta\ldots}(x,0) + \ldots \Big]
\notag\\[5pt]
&\quad\quad
- \Big[ F{^\gamma}_{\beta\mu}(x,0)\, T{^{\alpha\ldots}}_{\gamma\ldots}(x,0) + \ldots \Big]
\notag\\[8pt]
&= \frac{\partial}{\partial x^\mu}\, T{^{\alpha\ldots}}_{\beta\ldots}(x,0)
+ \Big[ F{^\alpha}_{\gamma\mu}(x,0)\, T{^{\gamma\ldots}}_{\beta\ldots}(x,0) + \ldots \Big]
\notag\\[5pt]
&\quad\quad
- \Big[ F{^\gamma}_{\beta\mu}(x,0)\, T{^{\alpha\ldots}}_{\gamma\ldots}(x,0) + \ldots \Big]
\notag\\[8pt]
&= D_\mu\, T{^{\alpha\ldots}}_{\beta\ldots}(x,0) \,.
\label{eqn:NablahtoD}
\end{align}
The term that involves the non-linear connection $N{^\lambda}_\mu$ drops as long as $\big( \frac{\partial}{\partial y^\lambda}\, T{^{\alpha\ldots}}_{\beta\ldots} \big)\bigr|_{y=0}$ is well-defined, which follows from the analyticity of the d-tensor $T{^{\alpha\ldots}}_{\beta\ldots} (x,y)$ at $y=0$. This allows us to write the last line above, in which we view $F{^\alpha}_{\beta\gamma}(x,0)$ as the Christoffel symbols for an affine connection $D$ on $\mathcal{M}$. We can iterate \cref{eqn:NablahtoD} to show that this applies to higher-order horizontal covariant derivatives as well:
\begin{equation}
T{^{\alpha\ldots}}_{\beta\ldots / \mu_1 \ldots \mu_n} (x,0) = D_{\mu_n} \cdots D_{\mu_1}\, T{^{\alpha\ldots}}_{\beta\ldots}(x,0) \,.
\end{equation}
Therefore, the horizontal geometry on the null section is completely governed by the connection $D$ (and the fundamental tensor $\mathfrak{g}_{\alpha\beta}(x, 0) = g_{\alpha\beta}(x)$).

\begin{table}[tbp]
\renewcommand{\arraystretch}{1.5}
\setlength{\arrayrulewidth}{0.2mm}
\setlength{\tabcolsep}{0.8em}
\centering
\begin{tabular}{cccc}
    \toprule
    Connections & $F\indices{^\alpha_\beta_\gamma}(x, 0)$ & $\bar{\mathcal{R}}_{\alpha\beta\gamma\delta}$ & $P\indices{^\alpha_\beta_\gamma}(x, 0)$ \\
    \midrule
    Levi-Civita & $\Gamma\indices{^\alpha_\beta_\gamma}(x)$ & $\bar{R}_{\alpha\beta\gamma\delta}$ & --- \\
    Cartan & $\Gamma\indices{^\alpha_\beta_\gamma}(x)$ & $\bar{R}_{\alpha\beta\gamma\delta}$ & $-3 c\indices{^\alpha^\delta_\beta_\gamma}(x) V_{,\delta}(x)$ \\
    Berwald & $\Gamma\indices{^\alpha_\beta_\gamma}(x) + 3 c\indices{^\alpha^\delta_\beta_\gamma}(x) V_{,\delta}(x)$ & $\bar{R}_{\alpha\beta\gamma\delta} + 6 \bar{c}\indices{_{\alpha\beta [\delta}^\epsilon} \bar{V}_{,\gamma] \epsilon}$ & 0 \\
    \bottomrule
\end{tabular}
\caption{A comparison of geometric quantities obtained from the Levi-Civita connection on $\mathcal{M}$, with those on the null section of $T\mathcal{M}$ obtained from common choices of $N$-linear connection. $\Gamma{^\alpha}_{\beta\gamma}$ and $R_{\alpha\beta\gamma\delta}$ are defined in terms of $g_{\alpha\beta}$ as in \cref{eqn:LeviCivitaConnection,eqn:RiemannTensor}. For brevity, we show the hh-curvature $\bar{\mathcal{R}}_{\alpha\beta\gamma\delta}$ only at the vacuum. The v-Christoffel symbols $C\indices{^\alpha_\beta_\gamma}$ and all other curvature and torsion components vanish on the null section.}
\label{tab:geomslit}
\end{table}

Since $F{^\alpha}_{\beta\gamma}(x,0)$ determines the horizontal geometry of the null section of $T\mathcal{M}$, it is illuminating to compare its value under common choices of $N$-linear connections with the Levi-Civita connection $\Gamma{^\alpha}_{\beta\gamma}(x)$ on $\mathcal{M}$. This is done in \cref{tab:geomslit} for the field theory Lagrangian \cref{eqn:Lagwithc}.\footnote{We listed four different $N$-linear connections in \cref{tab:Connections}, but either choice of $C{^\alpha}_{\beta\gamma}$ vanishes on the null section. In fact, the choice of $C{^\alpha}_{\beta\gamma}$ does not matter over the course of this paper, even when computing the vertical covariant derivatives in \cref{eqn:c2kasTorsion} later, so two names --- Cartan and Berwald --- have been selected without prejudice for simplicity.} We see that the horizontal geometry of the Cartan connection reproduces the Riemannian geometry on $\mathcal{M}$, but extra information appears in the vertical torsion component $P{^\alpha}_{\beta\gamma}$ in the form of higher-point Wilson coefficients. Moreover, a special status is granted to the new Berwald connection in Lagrange space, whose h-Christoffel symbols are notably always the difference between $F{^\alpha}_{\beta\gamma}$ and $P{^\alpha}_{\beta\gamma}$.

The point $(\bar{x}, 0)$ on the null section where $\bar{x}$ minimizes $V(x)$ will be called the vacuum on $T\mathcal{M}$, in analogy with the case of Riemannian geometry on $\mathcal{M}$. All quantities at this point are denoted with a bar, and in this paper we will study them in detail. While restricting to the vacuum may not retain all the physics inscribed on $T\mathcal{M}$ at $y \neq 0$, it already yields simple explicit expressions from which new and interesting lessons can be learned.

\section{Scattering Amplitudes as Horizontal Geometry}
\label{sec:horgeom}

We have seen how the Lagrange geometry of $\mathcal{M}$ constructed in \cref{sec:lagrange} is a generalization of Riemannian geometry --- the purely horizontal aspects of the Cartan connection match the Levi-Civita connection on the null section. We also know that the Riemannian structure of $\mathcal{M}$ encapsulates tree-level scattering amplitudes in the geometry at the vacuum. It follows at once that the Lagrange construction does the same via the horizontal geometry at its analogous vacuum. In fact, the explicit result \cref{eqn:AnptLC} from Riemannian geometry for a two-derivative Lagrangian carries over essentially verbatim whether we adopt the Cartan or Berwald connection:
\begin{align}
\left ( \prod_{i=1}^n \sqrt{2 \bar{\mathfrak{g}}_{\alpha_i \alpha_i}} \right ) \Amp_n &= \bar{V}_{/(\alpha_1 \ldots \alpha_n)} - 2\, \frac{n-3}{n-1} \, \sum_{i<j} s_{ij} \left [ \bar{\mathcal{R}}_{\alpha_i (\alpha_1 \alpha_2 | \alpha_j / | \alpha_3 \ldots \alpha_n)} + \mathcal{O}(\bar{\mathcal{R}}^2) \right ]
\notag\\[5pt]
&\quad
+ \text{ factorizable pieces} \,.
\label{eqn:AnptBerwaldnoc}
\end{align}
This is because they both agree with the Levi-Civita connection when higher-point Wilson coefficients vanish (see \cref{tab:geomslit}). We see in \cref{eqn:AnptBerwaldnoc} that order by order in kinematics, the coefficients have been written in terms of geometric quantities of the Lagrange space at the vacuum.\footnote{The masses-squared in the propagators of factorizable pieces can always be written as ratios of second-order h-covariant derivatives of $V$ to the fundamental tensor at the vacuum, if one demands a technically fully covariant expression for the scattering amplitude. See \cref{eq:massesToPotential}.} Notably, aside from mass factors, the order-$s$ piece is again strictly determined by the fundamental tensor without the need for an additional structure given by the (d-)scalar $V$. Thus, it is fair to say that the Lagrange (and Riemannian) construction encodes part of the scattering physics in its intrinsic geometry alone.

What is new and hence more powerful about the horizontal geometry of the Lagrange space is the fact that the Berwald connection on the null section differs from the Levi-Civita connection on $\mathcal{M}$ for a generic Lagrangian in \cref{eqn:Lagwithc} that has higher-derivative interactions. For simplicity, in this section let us focus on an example Lagrangian with up to four-derivative interactions:
\begin{equation}
\mathcal{L}(x, y) = V(x) + g_{\alpha\beta}(x)\, y^\alpha y^\beta + c_{\alpha\beta\gamma\delta}(x)\, y^\alpha y^\beta y^\gamma y^\delta + \mathcal{O}(y^6) \,.
\label{eqn:simpleLag}
\end{equation}
The question then is how the scattering physics from these four-derivative interactions can be interpreted using the horizontal geometry of the Berwald connection --- we would like to find the generalization of \cref{eqn:AnptBerwaldnoc} when $c_{\alpha\beta\gamma\delta}(x)$ is turned on. In what follows, we will see that the physics encoded in $c_{\alpha\beta\gamma\delta}(x)$ can be described using the horizontal geometry of an $N$-linear connection $\nabla$ on the Lagrange space. While any $N$-linear connection $\nabla$ corresponding to a torsion-free affine connection $D$ on $\mathcal{M}$ will suffice, the choice of connection determines whether the physics is attributed to the intrinsic geometry that follows from $\nabla$, or to additional contributions from the d-tensors $V$ and $c$.

\subsection{Encoding Amplitudes in the Geometry of General Connections}

We have learned in the previous section that the null-section horizontal geometry of an $N$-linear connection $\nabla$ on $T\mathcal{M}$ is equivalent to that of an affine connection $D$ on $\mathcal{M}$. The latter is moreover torsion-free for both the Cartan and Berwald connections:
\begin{equation}
F{^\alpha}_{\beta\gamma}(x,0) = F{^\alpha}_{\gamma\beta}(x,0) \,,
\end{equation}
as can be verified from \cref{tab:geomslit}.\footnote{Note that torsion-freeness on $\mathcal{M}$ is equivalent to the vanishing of the (h)h-torsion $T\indices{^\alpha_{\beta\gamma}}$ on the null section of $T\mathcal{M}$, given in \cref{eqn:Torsion}. This is not to be confused with other torsion components on $T\mathcal{M}$ like the (v)hv-torsion $P{^\alpha}_{\beta\gamma}$.} 
So let us tackle the same problem but for $\mathcal{M}$ endowed with an arbitrary torsion-free connection $D$ whose Christoffel symbols $F{^\alpha}_{\beta\gamma}(x,0)$ are different from $\Gamma{^\alpha}_{\beta\gamma}$ in general, so that $\mathcal{M}$ is not necessarily Riemannian.

To find the generalization of \cref{eqn:AnptBerwaldnoc} that incorporates $c_{\alpha\beta\gamma\delta}(x)$, let us compute the tree-level amplitudes for the Lagrangian in \cref{eqn:simpleLag}. They can be constructed from the momentum space Feynman rules (again in diagonalized coordinates):
\begingroup
\allowdisplaybreaks
\begin{subequations}\label{eqn:FeynRule}
\begin{align}
\vcenter{\hbox{ \begin{tikzpicture} \begin{feynman}
        \vertex (a) at (-0.8,0) {$\alpha$};
        \vertex (b) at (0.8,0) {$\beta$};
        \diagram* {(a) -- (b)};
    \end{feynman} \end{tikzpicture}}}
&\quad=\;\;
\frac{i\, \bar{g}^{\alpha \beta}}{2\, (p^2 - m_\alpha^2)} \,, \label{eqn:FeynProp} \\[10pt]
\vcenter{\hbox{ \begin{tikzpicture} \begin{feynman}
        \vertex (m) at (0,0);
        \vertex (a1) at (1.5,0) {$\alpha_1$};
        \vertex (a2) at (1.3,0.8) {$\alpha_2$};
        \vertex (ar) at (1.3,-0.8) {$\alpha_r$};
        \diagram* {
        (a1) -- (m) -- (a2),
        (ar) -- (m)};
        \draw [dotted,thick,domain=45:315] plot ({cos(\x)}, {sin(\x)});
    \end{feynman} \end{tikzpicture}}}
&\quad=\;\;
i \X_{\alpha_1 \ldots \alpha_r} \,, \label{eqn:FeynVert}
\end{align}
\end{subequations}
\endgroup
where the $r$-point vertex function $\X_{\alpha_1 \ldots \alpha_r}$ is
\begin{align}
\X_{1\ldots r} &= \bar{V}_{, \,\ldots} - 2 \sum_{i < j} \left(p_i \cdot p_j\right) \bar{g}_{ij, \,\ldots} + 8 \sum_{i<j,\, i<k<l} \left(p_i \cdot p_j\right) \left(p_k \cdot p_l\right) \bar{c}_{ijkl, \,\ldots}
\notag\\[8pt]
&= \bar{V}_{, \,\ldots} 
- \sum_{i<j} s_{ij}\, \bar{g}_{ij, \,\ldots} 
+ (r-1) \sum_i p_i^2\, \bar{g}_{i (1, \,\ldots)}
+ 2 \sum_{i<j,\, i<k<l} s_{ij} s_{kl}\, \bar{c}_{ijkl, \,\ldots}
\notag\\[8pt]
&\quad
- 2\, (r-3) \sum_{i<j} s_{ij} \sum_{k \neq i,j} p_k^2\, \bar{c}_{ijk (1, \,\ldots)}
+ 2\, (r-2)(r-3) \sum_{i<j} p_i^2 p_j^2\, \bar{c}_{ij (12, \,\ldots)} \,.
\label{eqn:Xr}
\end{align}
The shorthand $\alpha_i \to i$ has been adopted for particle flavor indices for concision, and the ellipses represent indices from $\alpha_1$ to $\alpha_r$ that have not been explicitly written. The explicit indices $i,j,k$ and $l$ all need to be distinct.

Since each diagram is a tree, any momentum in the Feynman rules can be written as sums of external momenta and traded into Mandelstam variables; they form the kinematic portion of the amplitude and we need only worry about the geometric representation of the remaining factors. Any mass-squared factor can be properly rewritten using $\bar{V}$ and $\bar{g}$, whether it multiplies a tensor:
\begin{equation}
-2\, m_i^2\, \bar{T}_{\ldots i \ldots} = \bar{V}_{,ij}\, \bar{g}^{jk}\, \bar{T}_{\ldots k \ldots} \,,
\label{eq:massesToPotential}
\end{equation}
or stands alone: $- m_i^2 = \bar{g}^{ii} \, \bar{V}_{,ii} \, / \, 2$. Therefore, setting the kinematic portion aside, all remaining factors in the amplitudes are in the form of partial derivatives of $V$, $g$, and $c$ evaluated at the vacuum.

For a moment, let us adopt the geodesic normal coordinates of $D$ at the vacuum. These are guaranteed to exist and be sufficiently well-behaved~\cite{Iliev:2006et}.\footnote{One may at first glance object that normal coordinates are typically not well-behaved for Finsler space, a special class of Lagrange spaces. But here, we are constructing normal coordinates on $\mathcal{M}$ only and not $T\mathcal{M}$. The results derived in these coordinates apply to the field theory Lagrange space because we are for now only interested in the horizontal geometry of the null section.} In \cref{app:Vgccov}, we show that in such coordinates, one can replace partial derivatives with covariant ones:
\begin{subequations}\label{eqn:VgcNormal}
\begin{align}
\bar{V}_{,\,\ldots} &\quad\longrightarrow\quad
\bar{V}_{/(\ldots)} \,, \\[5pt]
\bar{g}_{ij,\,\ldots} &\quad\longrightarrow\quad
\bar{g}_{ij/ (\ldots)} + \frac{r-3}{r-1}\, \Big[ \bar{\mathcal{R}}_{i(\ldots|j/|\ldots)} + \bar{\mathcal{R}}_{j(\ldots|i/|\ldots)} \Big]
+ \mathcal{O}(g_/ \mathcal{R},\, \mathcal{R}^2) \,, \\[5pt]
\bar{c}_{ijkl, \,\ldots} &\quad\longrightarrow\quad
\bar{c}_{ijkl / (\ldots)} + \mathcal{O}(c \mathcal{R}) \,,
\end{align}
\end{subequations}
Here the symbols $/$ and $\mathcal{R}$, originally meant for the connection $\nabla$ on $T\mathcal{M}$, have been reused for the connection $D$ on $\mathcal{M}$, due to their equivalence explained in \cref{sec:lagrange}. Note that $\mathcal{R}_{i(\ldots|j/|\ldots)}$ is not automatically symmetric between $i$ and $j$, as the connection is not necessarily Riemannian. The above replacements are valid only when the partial derivatives on the left-hand side are in normal coordinates. But we know for a fact that scattering amplitudes are covariant, so the ensuing tensorial expression obtained for the overall amplitude must hold for any coordinates, even if the intermediate factors do not.

Before we proceed to enumerate all diagrams and substitute covariant tensors for all non-kinematic quantities, let us consider how each diagram can be decomposed into a) ``atomic'' pieces that are polynomial in kinematic invariants, and b) pieces that can be recursively constructed by stitching expressions for lower-point on-shell amplitudes together using propagator factors. This allows a comparison of the ``atomic'' --- i.e., non-recursively constructible --- pieces to the previous expressions \cref{eqn:AnptLC,eqn:AnptBerwaldnoc}, which are a limiting case with no four-derivative interaction and essentially Riemannian geometry.\footnote{Note the difference between ``recursively constructible'' here and ``factorizable'' in the two-derivative Riemannian case. This distinction accommodates for the possibility that higher-order polynomials in Mandelstam invariants, arising from four- and higher-derivative terms in the Lagrangian, may cancel propagator poles in the denominators of recursively constructed pieces.} To this end, consider the vertex function $\X_r$, in which each leg has a particle flavor $\alpha_i$ and an ingoing momentum $p_i$. When the vertex is part of an amplitude diagram, the momentum of an external leg will be on-shell, i.e. $p_i^2 = m_i^2$, while that of an internal line (propagator) may be off-shell. Splitting each internal momentum into ``on-shell'' and ``off-shell'' pieces:
\begin{equation}
p_i^2 = m_i^2 + \left( p_i^2 - m_i^2 \right) \,,
\label{eqn:p2split}
\end{equation}
allows us to decompose the vertex function as
\begin{equation}
\X_{1 \ldots r} = \V_{1 \ldots r}
+ \sum_i \V_{1 \ldots \underline{i} \ldots r} \left( p_i^2 - m_i^2 \right)
+ \sum_{i<j} \V_{1 \ldots \underline{i} \ldots \underline{j} \ldots r} \left( p_i^2 - m_i^2 \right) \left( p_j^2 - m_j^2 \right) \,.
\label{eqn:XrDecompose}
\end{equation}
An underline indicates a leg that is ``off-shell''. For $r\ge 4$, the coefficients can be directly read off \cref{eqn:Xr} as
\begin{subequations}\label{eqn:Vr}
\begin{align}
\V_{1 \ldots r} &=
\bar{V}_{, \,\ldots}
+ (r-1) \sum_i m_i^2\, \bar{g}_{i (1, \,\ldots)}
+ 2\, (r-2)(r-3) \sum_{i<j} m_i^2 m_j^2\, \bar{c}_{ij (12, \,\ldots)}
\notag\\[5pt]
&\quad
- \sum_{i<j} s_{ij}\, \bar{g}_{ij, \,\ldots} 
- 2\, (r-3) \sum_{i<j} s_{ij} \sum_{k \neq i,j} m_k^2\, \bar{c}_{ijk (1, \,\ldots)}
\notag\\[5pt]
&\quad
+ 2 \sum_{i<j,\, i<k<l} s_{ij} s_{kl}\, \bar{c}_{ijkl, \,\ldots} \,, \label{eqn:Vr0} \\[10pt]
\V_{1 \ldots \underline{i} \ldots r} &= (r-1)\, \bar{g}_{i(1, \,\ldots)}
+ 2\, (r-2)(r-3)\, \sum_{j\ne i} m_j^2\, \bar{c}_{ij(12, \,\ldots)}
\notag\\[5pt]
&\quad
- 2\, (r-3) \sum_{j<k,j\ne i,k\ne i} s_{jk}\, \bar{c}_{ijk(1, \,\ldots)} \,, \label{eqn:Vr1} \\[10pt]
\V_{1 \ldots \underline{i} \ldots \underline{j} \ldots r} &= 2\,(r-2)(r-3)\, \bar{c}_{ij(12, \,\ldots)} \,. \label{eqn:Vr2}
\end{align}
\end{subequations}
Note that per our definition, the symbol $\V$ is fully symmetric under indices that are not underlined, while the positions of the underlined indices are not meaningful, so any ordering of the indices in $\V$ refers to the same quantity.

Meanwhile, $r=3$ is a special case --- although \cref{eqn:Xr} still holds, momentum conservation relates the Mandelstam variables $s_{ij}$ to the momentum squared $p_i^2$, such as in
\begin{equation}
s_{12} = \left( p_1 + p_2 \right)^2 = p_3^2 \,.
\end{equation}
In this case, there is not a unique way to separate the expression into the $s_{ij}$ and $p_i^2$ terms. We therefore combine these two types of terms:
\begin{align}
X_{123} &= \bar{V}_{, 123}
- \sum_i s_{jk}\, \bar{g}_{jk, i} 
+ 2 \sum_i p_i^2\, \bar{g}_{i (j, k)}
\notag\\[5pt]
&= \bar{V}_{, 123}
+ \sum_i p_i^2\, \left( \bar{g}_{ij, k} + \bar{g}_{ik, j} - \bar{g}_{jk, i} \right) \,,
\end{align}
leading to different expressions from \cref{eqn:Vr0,eqn:Vr1}:
\begin{subequations}\label{eqn:V3}
\begin{align}
\V_{123} &= \bar{V}_{, 123} + \sum_i m_i^2 \left( \bar{g}_{ij,k} + \bar{g}_{ik,j} - \bar{g}_{jk,i} \right) \,, \label{eqn:V30} \\[5pt]
\V_{\underline{i}\hspace{1pt}j\hspace{1pt}k} &= \bar{g}_{ij,k} + \bar{g}_{ik,j} - \bar{g}_{jk,i} \,, \label{eqn:V31} \\[5pt]
\V_{\underline{i\hspace{1pt}j}\hspace{1pt}k} &= 0 \,, \label{eqn:V32}
\end{align}
\end{subequations}
with $i,j$ and $k$ all distinct as before.

Using the replacements in \cref{eqn:VgcNormal}, one can rewrite the $\V$ factors as covariant tensors. For $r\ge 4$, the expressions in \cref{eqn:Vr} become
\begin{subequations}\label{eqn:VrNormal}
\begin{align}
\V_{1 \ldots r} &\;\longrightarrow\;
\bar{V}_{/(\ldots)} - \frac{r-1}{2} \sum_i \bar{V}_{/i \gamma}\, \bar{g}\indices{^\gamma_{(1/\ldots)}}
- \frac{r-3}{2} \sum_i \bar{V}_{/i \gamma}\, \bar{g}^{\gamma \delta} \bar{\mathcal{R}}_{(1\cdots|\delta/|\cdots)}
\notag\\[5pt]
&\;\;\quad\quad
+ \frac{(r-2)(r-3)}{2} \sum_{i<j} \bar{V}_{/i \gamma} \bar{V}_{/j \delta}\, \bar{c}\indices{^\gamma^\delta_{(12 / \ldots)}}
+ (r-3) \sum_{i<j} s_{ij} \sum_{k \neq i,j} \bar{V}_{/k \gamma}\, \bar{c}\indices{^\gamma_{ij (1/\ldots)}}
\notag\\[5pt]
&\;\;\quad\quad
- \sum_{i<j} s_{ij} \bigg\{
\bar{g}_{ij/(\ldots)} + \frac{r-3}{r-1}\, \Big[ \bar{\mathcal{R}}_{i(\ldots|j/|\ldots)}
+ \bar{\mathcal{R}}_{j(\ldots|i/|\ldots)} \Big] \bigg\}
\notag\\[5pt]
&\;\;\quad\quad
+ 2 \sum_{i<j,\, i<k<l} s_{ij} s_{kl}\, \bar{c}_{ijkl/(\ldots)} + \mathcal{O}(g_{/} \mathcal{R},\, c\mathcal{R},\, \mathcal{R}^2) \,, \\[10pt]
\V_{1\ldots \underline{i}\ldots r} &\;\longrightarrow\;
(r-1)\, \bar{g}_{i(1/\ldots)} + (r-3)\, \bar{\mathcal{R}}_{(123|i/| \ldots)}
- (r-2)(r-3) \sum_{j\ne i} \bar{V}_{/j \gamma}\, \bar{c}\indices{^\gamma_{i(12/\ldots)}}
\notag\\[5pt]
&\;\;\quad\quad
- 2\, (r-3) \sum_{j<k, j\ne i,k\ne i} s_{jk} \, \bar{c}_{ijk(1/\ldots)}
+ \mathcal{O}(g_{/} \mathcal{R},\, c\mathcal{R},\, \mathcal{R}^2) \,, \\[8pt]
\V_{1\ldots \underline{i}\ldots \underline{j}\ldots r} &\;\longrightarrow\;
2\, (r-2)(r-3)\, \bar{c}_{ij(12/ \ldots)} + \mathcal{O}(g_{/} \mathcal{R},\, c\mathcal{R},\, \mathcal{R}^2) \,,
\end{align}
\end{subequations}
and for $r=3$, \cref{eqn:V3} becomes
\begin{subequations}\label{eqn:V3Normal}
\begin{align}
\V_{123} &\;\longrightarrow\;
\bar{V}_{/(123)} - \frac12 \sum_i \bar{V}_{/i\gamma}\, \bar{g}^{\gamma \delta} \left( \bar{g}_{\delta j/k} + \bar{g}_{\delta k/j} - \bar{g}_{jk/\delta} \right) \,, \\[5pt]
\V_{\underline{i}\hspace{1pt}j\hspace{1pt}k} &\;\longrightarrow\;
\bar{g}_{ij/k} + \bar{g}_{ik/j} - \bar{g}_{jk/i} \,, \label{eqn:V31Normal} \\[8pt]
\V_{\underline{i\hspace{1pt}j}\hspace{1pt}k} &\;\longrightarrow\; 0 \,.
\end{align}
\end{subequations}

We are now ready to build up the full $n$-point scattering amplitude $\mathcal{A}_n$ from its constituents. As emphasized, our priority will be to collect the non-recursively constructible pieces. We begin with the $3$-point amplitude, which is simple as it receives contributions only from the contact diagram, with all the legs having on-shell momenta:
\begin{equation}
\left( \prod_{i=1}^3 \sqrt{2 \bar{g}_{ii}} \right) \Amp_3 = \X_{123} = \V_{123} \,.
\label{eq:A3}
\end{equation}
This is the only piece in the amplitude.

At $n=4$, propagator diagrams start to appear:
\begin{align}
\left( \prod_{i=1}^4 \sqrt{2 \bar{g}_{ii}} \right) \Amp_4 &= \X_{1234}
+ \frac12\, \binom{4}{2}\;\text{distinct perms. of}\;
\left[ - \frac12\, \X_{125}\, \frac{\bar{g}^{56}}{p_5^2 - m_5^2}\, \X_{346} \right] \,.
\label{eqn:A4X}
\end{align}
Within the square brackets, the minus sign collects the factors of $i$ from the propagator and the second vertex, and the factor $1/2$ arises from the propagator. Similar to \cref{eqn:A4ptLC}, the flavor indices ``$5$'' and ``$6$'' in the propagator (which are abbreviations for $\alpha_5$ and $\alpha_6$) are dummy indices summed over all field flavors, and we are working in coordinates such that the propagator matrix is diagonal. For the $4$-point amplitude, there are three distinct permutations (channels) of the term in the square bracket; each of them yields both recursively constructible and non-recursively constructible contributions. Recalling the decomposition in \cref{eqn:XrDecompose} of the $3$-point vertex function (with the $\V$ factors given in \cref{eqn:V3Normal}):
\begin{subequations}
\begin{align}
\X_{125} = \V_{125} + \V_{12\underline{5}} \left( p_5^2 - m_5^2 \right) \,, \\[5pt]
\X_{346} = \V_{346} + \V_{34\underline{6}} \left( p_6^2 - m_6^2 \right) \,,
\end{align}
\end{subequations}
we extract the non-recursively constructible pieces as follows:
\begin{align}
\X_{125}\, \frac{\bar{g}^{56}}{p_5^2 - m_5^2}\, \X_{346} &=
\V_{12\underline{5}}\, \V{^5}_{34} + \V_{34\underline{5}}\, \V{^5}_{12}
+ \sum_{\alpha_5} \V_{12\underline{5}}\, \V{^{\underline{5}}}_{34}
\left( s_{12} + \tfrac12\, \bar{g}^{55} \bar{V}_{/55} \right)
\notag\\[5pt]
&\quad
+ \,\V_{125}\, \frac{\bar{g}^{56}}{p_5^2 - m_5^2}\, \V_{346} 
\notag\\[10pt]
&=
\V_{12\underline{5}}\, \V{^5}_{34} + \V_{34\underline{5}}\, \V{^5}_{12}
+ \sum_{\alpha_5} \V_{12\underline{5}}\, \V{^{\underline{5}}}_{34}
\left( s_{12} + \tfrac12\, \bar{g}^{55} \bar{V}_{/55} \right)
\notag\\[5pt]
&\quad
+ \text{ recursively constructible pieces} \,.
\end{align}
where indices on the $\V$ factors have been raised and lowered using $\bar{g}$. We see that the last term takes the form of two (``atomic'') 3-point amplitudes, \cref{eq:A3}, multiplied by an explicit propagator factor. This is recursively constructible from lower-point on-shell amplitudes. Substituting this back in \cref{eqn:A4X} yields
\begin{align}
\left( \prod_{i=1}^4 \sqrt{2 \bar{g}_{ii}} \right) \Amp_4 &= \V_{1234}
+ \binom{4}{2}\;\text{distinct perms. of}\;
\left[ - \frac12\, \V_{12\underline{5}}\, \V{^5}_{34} \right]
\notag\\[5pt]
&\quad
+ \frac12\, \binom{4}{2}\;\text{distinct perms. of}\;
\left[ - \frac12 \sum_{\alpha_5} \V_{12\underline{5}}\, \V{^{\underline{5}}}_{34}
\Big( s_{12} + \tfrac12\, \bar{g}^{55} \bar{V}_{/55} \Big) \right]
\notag\\[5pt]
&\quad
+ \text{ recursively constructible pieces} \,.
\label{eqn:A4V}
\end{align}
We see how the $\V$ factors help organize the expressions for the non-recursively constructible pieces.

Similarly, non-recursively constructible pieces for higher-point amplitudes $\Amp_n$ can be systematically enumerated using the bookkeeping via $\V$. In general, each diagram generates a string of $\V$ factors following the rules below:
\begin{itemize}
\item The overall constant factor is $-1/2$ to the power of the number of propagators.
\item Each propagator yields a dummy flavor index $\alpha_i$ (abbreviated as $i$, like the ``$5$'' in \cref{eqn:A4V}) that splices a contracted pair of $\V$ factors.
\item At least one in each pair of contracted dummy flavor indices $i$ needs to be underlined, i.e. all propagators should be canceled. Otherwise, the term is a recursively constructible piece. If both dummy indices $i$ at a propagator are underlined (such as in the second line of \cref{eqn:A4V}), there will be an extra factor of $p_i^2 - m_i^2$, which can be rewritten using Mandelstam variables and
\begin{equation}
- m_i^2 = \frac12\, \bar{g}^{ii}\, \bar{V}_{/ii} \,.
\end{equation}
In this case, we will explicitly write out the sum for the dummy flavor index $\alpha_i$ to avoid confusion.
\item Similar terms are obtained by permutations of the uncontracted indices in distinct ways, watching out for the symmetries of the string of $\V$ factors.
\end{itemize}
For the sake of exposition, let us reduce the length of our expressions by assuming from now on that the connection $D$ on $\mathcal{M}$ satisfies $\bar{g}_{\alpha \beta / \gamma} = 0$,\footnote{Both the Cartan and Berwald connections correspond to connections on $\mathcal{M}$ that satisfy this condition.} so that the $3$-point vertex function only has an on-shell piece, i.e. $\V_{\underline{i}\hspace{1pt}j\hspace{1pt}k}=0$ (see \cref{eqn:V31Normal}) in addition to $\V_{\underline{i\hspace{1pt}j}\hspace{1pt}k}=0$. Under this assumption, the two terms from the propagator diagram in \cref{eqn:A4V} are zero, and the first non-recursively constructible contribution from propagator diagrams arises at $n=5$ instead:
\begin{align}
\left( \prod_{i=1}^5 \sqrt{2 \bar{g}_{ii}} \right) \Amp_5 &= \V_{12345}
+ \binom{5}{2}\;\text{distinct perms. of}\;
\left[ - \frac12\, \V_{123 \underline{6}}\, \V\indices{^6_{45}} \right]
\notag\\[5pt]
&\quad
+ \text{ recursively constructible pieces} \,.
\end{align}
Higher-point amplitudes follow in the same way. For example, $\Amp_6$ is given by
\begin{align}
\left( \prod_{i=1}^6 \sqrt{2 \bar{g}_{ii}} \right) \Amp_6 &= \V_{123456}
+ \binom{6}{2}\;\text{distinct perms. of}\;
\left[ -\frac12\, \V_{1234 \underline{7}}\, \V\indices{^7_{56}} \right]
\notag\\[5pt]
&\quad
+ \binom{6}{3}\;\text{distinct perms. of}\;
\left[ -\frac12\, \V_{123\underline{7}}\, \V\indices{^7_{456}} \right]
\notag\\[5pt]
&\quad
+ \frac12\, \binom{6}{3}\;\text{distinct perms. of}\;
\left[ -\frac12\, \sum_{\alpha_7} \V_{123\underline{7}}\, \V\indices{^{\underline{7}}_{456}} \Big( s_{123} + \tfrac12\, \bar{g}^{77} {\bar{V}_{/77}} \Big) \right]
\notag\\[5pt]
&\quad
+ \frac12\, \binom{6}{2,2}\;\text{distinct perms. of}\;
\left[ +\frac14\, \V_{127}\, \V\indices{^{\underline{7}}_{34 \underline{8}}}\, \V\indices{^8_{56}} \right]
\notag\\[5pt]
&\quad
+ \text{ recursively constructible pieces} \,. 
\end{align}
Combined with the explicit covariant expressions of the $\V$ factors given in \cref{eqn:VrNormal,eqn:V3Normal}, these comprise tensorial expressions for $\Amp_n$, valid in all coordinates since scattering amplitudes are covariant quantities. Translating these results to field theory Lagrange spaces is a simple matter of identifying $D$ with an $N$-linear connection $\nabla$ whose h-Christoffel symbols are $F$ and v-Christoffel symbols are zero at the null section. The tensors $V(x)$, $g(x)$, and $c(x)$ on $\mathcal{M}$ in \cref{eqn:VrNormal,eqn:V3Normal} are to be replaced by the d-tensors $V(x, y)$, $\mathfrak{g}(x, y)$, and $c(x, y)$ on $T\mathcal{M}$, the first and last being $y$-independent.

For the purpose of presentation, we have taken the example of a four-derivative Lagrangian of the form \cref{eqn:simpleLag}, and a connection that is h-metric-compatible at the vacuum to first order, i.e. $\bar{\mathfrak{g}}_{\alpha \beta / \gamma} = 0$. The relaxation of these assumptions is straightforward --- in particular, since we have opted to geometrize higher-order derivative physics using a separate d-tensor $c$, it need not be symmetric in its indices. However, the Berwald h-covariant derivative will only register the symmetric part.

A comparison with the Riemannian result \cref{eqn:AnptLC} under a two-derivative Lagrangian reveals what is new --- the introduction of higher-derivative interactions produces non-trivial physics at higher orders in momentum, appearing in the form of order-$s^2$ non-recursively constructible pieces.
Such physics can nevertheless be described using the horizontal geometry of an $N$-linear connection $\nabla$ on the Lagrange space. Note that any $N$-linear connection $\nabla$ that corresponds to a torsion-free affine connection $D$ on $\mathcal{M}$ is able to capture these higher-derivative interactions, as the expressions in \cref{eqn:VrNormal,eqn:V3Normal} are fully general. The choice of connection determines whether the physics is attributed to the intrinsic geometry that follows from $\nabla$, or to the additional structures as given by the d-tensors $V$ and $c$. This makes one wonder: are there privileged connections that inherently capture more higher-derivative physics through the intrinsic geometry?

\subsection{Higher-Derivative Physics as Expressed by Privileged Connections}
\label{subsec:HigherDerivative}

The first instance of higher-derivative physics occurs at $n = 4$. Taking $\bar{\mathfrak{g}}_{\alpha \beta / \gamma} = 0$ again for simplicity and substituting the $\V$ factors given in \cref{eqn:VrNormal,eqn:V3Normal} to the result in \cref{eqn:A4V} produce the covariant total scattering amplitude in its full glory:
\begin{align}
\left( \prod_{i=1}^4 \sqrt{2 \bar{\mathfrak{g}}_{ii}} \right) \Amp_4 &=
\frac12\, \binom{4}{2}\;\text{distinct perms. of}\;
\left[ - \frac12\, \bar{V}_{/(125)}\, \frac{\bar{\mathfrak{g}}^{56}}{s_{12} - m_5^2}\, \bar{V}_{/(346)} \right]
\notag\\[5pt]
&\quad
+ \bar{V}_{/(1234)} + \sum_i m_i^2 \left [3\, \bar{\mathfrak{g}}_{i(j/kl)} + \bar{\mathcal{R}}_{(jkl)i} \right ] - \sum_{i<j} s_{ij} \, \bar{\mathfrak{g}}_{ij/(kl)}
\notag\\[5pt]
&\quad
- \frac13 \sum_{i<j} s_{ij} \left[ \bar{\mathcal{R}}_{i(kl)j} + \bar{\mathcal{R}}_{j(kl)i} \right]
\notag\\[5pt]
&\quad
+ \left [4 \sum_{i<j} m_i^2 m_j^2 - 2 \sum_{i<j} s_{ij} \sum_{k \neq i,j} m_k^2 + 2 \sum_{i<j, i<k< l} s_{ij} s_{kl} \right ] \bar{c}_{1234} \,.
\label{eqn:A4Original}
\end{align}
We have written out the masses-squared in preference over $\bar{V}_/$ and $\bar{\mathfrak{g}}$, collected terms proportional to $\bar{c}$ at each order in $s$, and listed the recursively constructible pieces as well on the first line. As always the indices $i$, $j$, $k$, and $l$ are all distinct. Let us reinstate the assumption that $c$ is symmetric. The special status granted to the Berwald and Cartan connections in generic Lagrange spaces motivates us to study their affine combinations, with h-Christoffel symbols
\begin{equation}
F\indices{^\alpha_\beta_\gamma}(x,0) = \Gamma\indices{^\alpha_\beta_\gamma}(x) + 3 \left(1-b\right) c\indices{^\alpha^\delta_\beta_\gamma}(x)\, V_{,\delta}(x) \,,
\label{eqn:AffineCombination}
\end{equation}
on the null section for some constant $b$. The Berwald and Cartan connections correspond to $b=0$ and $b=1$ respectively. The geometric quantities at the vacuum for any $b$ can be recast into those under the Cartan or equivalently the Levi-Civita connection as
\begin{subequations}\label{eqn:htoLC}
\begin{align}
\bar{V}_{/ijk} &= \bar{V}_{;ijk} \,, \\[5pt]
\bar{V}_{/ijkl} &= \bar{V}_{;ijkl} - 12 \left(1-b\right) \left( m_i^2 + m_j^2 + 2m_k^2 \right) m_l^2\, \bar{c}_{ijkl} \,, \\[5pt]
\bar{\mathfrak{g}}_{ij/kl} &= 12 \left(1-b\right) m_l^2\, \bar{c}_{ijkl} \,, \\[5pt]
\bar{\mathcal{R}}_{ijkl} &= \bar{R}_{ijkl} - 6 \left(1-b\right) \left( m_k^2 - m_l^2 \right) \bar{c}_{ijkl} \,.
\end{align}
\end{subequations}

Our first objective is to verify as a sanity check that \cref{eqn:A4Original} indeed yields the same scattering amplitude for any $b$. The kinematic identities of four-point scattering:
\begin{subequations}
\begin{gather}
s \coloneqq s_{12} = s_{34} \,,\quad
t \coloneqq s_{13} = s_{24} \,,\quad
u \coloneqq s_{14} = s_{23} \,, \\[5pt]
s + t + u = \sum_i m_i^2 \,, \label{eqn:stusum}
\end{gather}
\end{subequations}
can be used to show that all the $b$-dependence is contained in the second line of \cref{eqn:A4Original}, and conspires to cancel out. A change in connection by varying $b$ amounts to a reshuffling of the contributions to the amplitude at different orders in momentum, which has to happen if \cref{eqn:A4Original} is to hold for all values of $b$.

Our second objective is to choose a connection that simplifies \cref{eqn:A4Original} so that a physically significant term is determined by a reduced number of geometric quantities. Observe that the last two terms on the second line of \cref{eqn:A4Original} are all proportional to $\bar{c}$ upon substituting \cref{eqn:htoLC}. Merging them with the last line in \cref{eqn:A4Original}, we obtain
\begin{align}
\left( \prod_{i=1}^4 \sqrt{2 \bar{\mathfrak{g}}_{ii}} \right) \Amp_4 &=
\frac12\, \binom{4}{2}\;\text{distinct perms. of}\;
\left[ - \frac12\, \bar{V}_{/(125)}\, \frac{\bar{\mathfrak{g}}^{56}}{s_{12} - m_5^2}\, \bar{V}_{/(346)} \right]
\notag\\[5pt]
&\quad
+ \bar{V}_{/(1234)}
- \frac13 \sum_{i<j} s_{ij} \left[ \bar{\mathcal{R}}_{i(kl)j} + \bar{\mathcal{R}}_{j(kl)i} \right]
\notag\\[5pt]
&\quad
+ 4 \left[ \left(3-2b\right) \sum_{i<j} m_i^2 m_j^2 - \left( st + tu + us \right) \right] \bar{c}_{1234} \,.
\label{eqn:A4Reordered}
\end{align}
Taking $b=3/2$ would then eliminate all order-$s^0$ and -$s^1$ terms proportional to $\bar{c}$.

Due to the on-shell kinematic relation in \cref{eqn:stusum}, there is an ambiguity in the $s_{ij}$ order counting within the four-point amplitude. We now see that the simple expression \cref{eqn:AnptBerwaldnoc} from Riemannian geometry, applied to $b = 3/2$, actually contains all physics up to order $s$ even with $c$ turned on, if the ambiguity is resolved by taking the order-$s^2$ piece to be $-4 \left( st + tu + us \right) \bar{c}_{1234}$. In this sense, the connection with $b = 3/2$ is privileged due to its minimal description of the physics up to order $s$, with $c$ embedded only in the connection and not additionally as a separate d-tensor on $T\mathcal{M}$. It should be remarked that the linear dependence of the Mandelstam variables allows some, but not arbitrary, freedom in re-organizing the physics of $c$, so neither the reshuffling by the choice of connection nor the privilege enjoyed by $b = 3/2$ is trivial.

At higher points, the analogs to the kinematic identities at $n = 4$ do not seem sufficient in achieving the second objective above, although they must necessarily attain the first objective. Nevertheless, the connections parameterized by $b$ still comprise a favored family. All of them satisfy $\bar{\mathfrak{g}}_{\alpha \beta / \gamma} = 0$ so that $\V_{\underline{i}\hspace{1pt}j\hspace{1pt}k} = 0$. In particular, the Cartan connection is metric-compatible and eliminates h-covariant derivatives of $\mathfrak{g}$ to all orders, as well as symmetrizations of the first two indices of $\mathcal{R}$.\footnote{If we additionally impose $c = 0$, the result \cref{eqn:AnptBerwaldnoc} at arbitrary points is recovered, with the contact term being the only non-recursively constructible piece since all $\V_{1 \ldots \underline{i} \ldots r}$ and $\V_{1 \ldots \underline{i} \ldots \underline{j} \ldots r}$ vanish.} Hence, it is fair to say that in the quest to encode more information from the Lagrangian in a single manifold, higher-derivative physics has been more simply framed in terms of geometry.

One might argue up to this point that Lagrange spaces have not really provided anything new. The horizontal geometry of the null section on $T\mathcal{M}$ is no more general than the geometry of $\mathcal{M}$ with a general symmetric affine connection $D$, and the results above can be reproduced so long as we agree to consider connections other than the Levi-Civita one. This, however, understates the role that Lagrange spaces play in the choice of connection. While endowing $\mathcal{M}$ with a well-chosen connection requires an inspired guess, Lagrange spaces provide a canonical route to, e.g., the Berwald connection and hence a natural setting for non-Riemannian geometry. Moreover, we will next see an instance when Lagrange geometry is truly essential: the vertical geometry on $T\mathcal{M}$.

\section{Physical Validity as Vertical Geometry}
\label{sec:vergeom}

Recall from \cref{sec:lagrange} that the Lagrange space of a generic field theory Lagrangian \cref{eqn:Lagwithc} features another non-vanishing geometric invariant apart from the hh-curvature $\mathcal{R}$. For the Cartan connection, and more generally any affine combination of the Berwald and Cartan connections given in \cref{eqn:AffineCombination}:\begin{equation}
F\indices{^\alpha_\beta_\gamma}(x,0) = \Gamma\indices{^\alpha_\beta_\gamma}(x) + 3 \left(1-b\right) c\indices{^\alpha^\delta_\beta_\gamma}(x)\, V_{,\delta}(x) \,,
\label{eqn:AffineCombination1}
\end{equation}
the (v)hv-torsion on the null section reads
\begin{equation}
P{^\alpha}_{\beta\gamma}(x,0) = F{^\alpha}_{\gamma\beta}(x,0) - \frac{\partial N{^\alpha}_{\beta}}{\partial y^\gamma}(x,0)
= -3\, b\, c{^{\alpha\delta}}_{\beta\gamma}(x)\, V_{,\delta}(x) \,.
\label{eqn:vhvP4pt}
\end{equation}
It vanishes at the vacuum, but its h-covariant derivative registers the four-point Wilson coefficient:
\begin{equation}
\bar{P}_{\alpha \beta \gamma / \delta} = 6\, b\, m_\delta^2\, \bar{c}_{\alpha \beta \gamma \delta} \,,
\label{eqn:c4asTorsion}
\end{equation}
where for simplicity diagonalized coordinates have been adopted as before. Unlike the horizontal geometry in \cref{sec:horgeom} which mixes the four-derivative interaction $c_{\alpha\beta\gamma\delta}(x)$ with the two-derivative term $g_{\alpha\beta}(x)$, this vertical torsion component isolates $c_{\alpha\beta\gamma\delta}(x)$ and thus captures any strictly higher-derivative physics.\footnote{We clarify that $c_{\alpha\beta\gamma\delta}(x)$ is taken to be symmetric here since the torsion $P$ arises from the intrinsic geometry of the Lagrange space.} As such, we expect the vertical geometry to reflect physical constraints on higher-derivative operators, such as positivity bounds, that follow from bedrock principles of the underlying quantum field theory. In this respect, physical theories correspond to a restricted class of vertical geometry.

\subsection{Four-Point Positivity Bounds as Sign Constraints}
\label{subsec:FourPointPositivity}

A prominent example of higher-derivative physics comes from positivity bounds, which state that certain higher-order Wilson coefficients must be positive for an effective field theory to be physical. In particular, for four-point amplitudes in the forward limit, crossing symmetry, analyticity and unitarity will constrain the sign of the order-$s^2$ term at small $s$~\cite{Adams:2006sv}, the reason for which we briefly review below.

For simplicity, let us assume all flavors of scalars have the same mass $m$, so that it is straightforward to construct the following superposed states:\footnote{The case with a general mass spectrum can be handled with some extra effort.}
\begin{subequations}\label{eqn:Superposition}
\begin{align}
\ket{1} &= \ket{3} = \sum_\alpha \sqrt{2 \bar{\mathfrak{g}}_{\alpha\alpha}}\, \rho_1^\alpha \ket{\phi_\alpha} \,, \\[5pt]
\ket{2} &= \ket{4} = \sum_\alpha \sqrt{2 \bar{\mathfrak{g}}_{\alpha\alpha}}\, \rho_2^\alpha \ket{\phi_\alpha} \,.
\end{align}
\end{subequations}
Here $\ket{\phi_\alpha}$ denotes a single-particle state of flavor $\alpha$, and the superposition coefficients are taken to be real numbers satisfying
\begin{equation}
\sum_\alpha 2\bar{\mathfrak{g}}_{\alpha\alpha} \left( \rho_1^\alpha \right)^2 = \sum_\alpha 2\bar{\mathfrak{g}}_{\alpha\alpha} \left( \rho_2^\alpha \right)^2 = 1 \,.
\end{equation}
Suppose these states scatter in the forward limit, with external momenta:
\begin{equation}
p_1 = -p_3 \,,\qquad
p_2 = -p_4 \,.
\end{equation}
The four-point amplitude $\Amp_4^\rho$ is then a function of the center-of-mass energy squared
\begin{equation}
s = \left( p_1 + p_2 \right)^2 = 2\, p_1 \cdot p_2 + 2\, m^2 \,.
\end{equation}

To obtain a positivity bound, consider an integral of the amplitude along a small counter-clockwise contour $\gamma$ around the origin:
\begin{equation}
I_4 \coloneqq \frac{1}{2 \pi i} \oint_\gamma \frac{ds}{s^3}\, \Amp_4^\rho (s) \,.
\label{eqn:I4def}
\end{equation}
We assume a small analytic region for $\Amp_4^\rho (s)$ around $s=0$ through which the contour passes.\footnote{This can be guaranteed, for instance, whenever $m^2>0$ by displacing the argument of the amplitude by an infinitesimal (and inconsequential) positive amount.} The integral picks out the coefficient of the order-$s^2$ term in $\Amp_4^\rho(s)$ at low energies. This energy regime is well described by the EFT. Assuming, for simplicity, sufficiently weak coupling that makes loop corrections negligible, we can use the tree-level result in \cref{eqn:A4Original} for $\Amp_4$, whose order-$s^2$ piece in the forward limit is
\begin{equation}
\left( \prod_{i=1}^4 \sqrt{2 \bar{\mathfrak{g}}_{\alpha_i\alpha_i}} \right) \Amp_4 = 4s^2\, \bar{c}_{\alpha_1 \alpha_2 \alpha_3 \alpha_4} + \mathcal{O}(s) \,.
\label{eqn:A4Forward}
\end{equation}
Dressing it up with the superposition in \cref{eqn:Superposition}, we get
\begin{equation}
\Amp_4^\rho = 4s^2\, \bar{c}_{\alpha_1 \alpha_2 \alpha_3 \alpha_4}\, \rho_1^{\alpha_1} \rho_2^{\alpha_2} \rho_1^{\alpha_3} \rho_2^{\alpha_4} + \mathcal{O}(s) \,,
\label{eqn:A4rho}
\end{equation}
and therefore
\begin{equation}
I_4 = 4\, \bar{c}_{\alpha_1 \alpha_2 \alpha_3 \alpha_4}\, \rho_1^{\alpha_1} \rho_2^{\alpha_2} \rho_1^{\alpha_3} \rho_2^{\alpha_4} \,.
\label{eqn:I41}
\end{equation}

On the other hand, let us invoke the fact that $\Amp_4^\rho(s)$ is everywhere analytic on the complex $s$ plane, except for poles and branch cuts on the real $s$ axis. We can deform $\gamma$ into another contour $\gamma'$ running just above and below the positive and negative parts of the real $s$ axis, combined with a boundary contour at infinity.
This does not change the value of the contour integral due to Cauchy's theorem. The boundary integral at infinity vanishes due to the Froissart-Martin bound on the UV completion~\cite{Chaichian:1987zt}, leaving us with
\begin{equation}
I_4 = \frac{1}{2 \pi i} \left ( \int_{-\infty}^{0} + \int_{0}^{\infty} \right ) \frac{ds}{s^3}\, \text{disc} \, \Amp_4^\rho(s) \,,
\label{eqn:I42}
\end{equation}
where the discontinuity across the real axis is defined as
\begin{equation}
\text{disc} \, \Amp_4^\rho(s) = \Amp_4^\rho(s + i\epsilon) - \Amp_4^\rho(s - i\epsilon) \,.
\end{equation}
To proceed, we make use of several general properties of the scattering amplitude in the forward limit:
\begin{enumerate}
\item Crossing symmetry under the exchange of the two external legs
\begin{equation}
\ket{2} \leftrightarrow \ket{4} \,,
\end{equation}
implies that $\Amp_4^\rho(s)$ is invariant under $p_2 \rightarrow -p_2$, namely
\begin{equation}
\Amp_4^\rho(-s) = \Amp_4^\rho(s + 4m^2) \,.
\end{equation}
This allows us to merge the two branches of the integral in \cref{eqn:I42} to get
\begin{equation}
I_4 = \frac{1}{2 \pi i} \int_{0}^{\infty} \frac{ds}{s^3}\, \Big[ \text{disc}\, \Amp_4^\rho(s) + \text{disc}\, \Amp_4^\rho(s+4m^2) \Big] \,.
\label{eqn:I43}
\end{equation}
\item Hermitian analyticity~\cite{Miramontes:1999gd} implies the Schwarz reflection principle
\begin{equation}
\Amp_4^\rho(s^*) = \left[ \Amp_4^\rho(s) \right]^* \,,
\end{equation}
which then gives
\begin{equation}
\text{disc}\, \Amp_4^\rho(s) = \Amp_4^\rho(s + i\epsilon) - \left[ \Amp_4^\rho(s + i\epsilon) \right]^* = 2 i \, \text{Im}\, \Amp_4^\rho(s) \,.
\end{equation}
Applying this to the expression in \cref{eqn:I43}, we get
\begin{equation}
I_4 = \frac{1}{\pi} \int_{0}^{\infty} \frac{ds}{s^3}\, \Big[ \text{Im}\, \Amp_4^\rho(s) + \text{Im}\, \Amp_4^\rho(s+4m^2) \Big] \,.
\label{eqn:I44}
\end{equation}
\item Unitarity in the form of the optical theorem implies that
\begin{equation}
\text{Im}\, \Amp_4^\rho(s) \ge 0
\,,
\end{equation}
for $s$ real. This means the expression in \cref{eqn:I44} must be non-negative, $I_4 \ge 0$.
\end{enumerate}

Applying the bound $I_4 \ge 0$ on the deformed integral to the original integral in the low-energy $s$ region, we get a positivity bound on the four-derivative Wilson coefficient of the EFT:
\begin{equation}
\bar{c}_{\alpha_1 \alpha_2 \alpha_3 \alpha_4}\, \rho_1^{\alpha_1} \rho_2^{\alpha_2} \rho_1^{\alpha_3} \rho_2^{\alpha_4} \ge 0 \,,
\end{equation}
for any choice of $\rho_1$ and $\rho_2$~\cite{Remmen:2019cyz}. This can be reinterpreted as a positivity bound on the (v)hv-torsion:
\begin{equation}
\bar{P}_{\alpha_1 \alpha_2 \alpha_3 / \alpha_4}\, \rho_1^{\alpha_1} \rho_2^{\alpha_2} \rho_1^{\alpha_3} \rho_2^{\alpha_4} \ge 0 \,.
\label{eqn:P4bound}
\end{equation}
if $b > 0$, which includes the Cartan connection case and the $b = 3/2$ connection advocated in \cref{subsec:HigherDerivative}. In other words, the h-covariant derivative of the (v)hv-torsion at the vacuum is a positive semi-definite biquadratic form. Notably, each diagonal component $\bar{P}_{\alpha_1 \alpha_2 \alpha_1 / \alpha_2}$ must be non-negative. In this sense, we can say that the (v)hv-torsion is zero but increasing at the vacuum.

The geometric interpretation of the constraint is as follows. Imagine that an observer travels along a path that starts horizontally at the vacuum. The observer carries a horizontal rod that points in the direction to move, and a vertical reference rod that is parallel-transported. After a while, the vertical rod will begin to rotate relative to the horizontal one, such that its movement is also vertical and aligned with the direction of the vertical rod, in the sense that their inner product with respect to the fundamental tensor is positive.

It deserves emphasis that the geometry of the manifold $\mathcal{M}$ does not typically allow us to isolate the four-derivative term $c_{\alpha\beta\gamma\delta}(x)$. There are two main geometric invariants arising from a connection on $\mathcal{M}$ --- curvature and torsion. Were the symmetric coefficient $c_{\alpha\beta\gamma\delta}(x)$ to appear in an invariant, it necessarily enters the curvature and is hence combined with the standard Riemannian part from $g_{\alpha\beta}(x)$. Meanwhile, the geometry of $T\mathcal{M}$ grants a special status to the Berwald connection so that its difference with a given $N$-linear connection appears in a vertical torsion component, capturing $c_{\alpha\beta\gamma\delta}(x)$ alone despite its symmetry. The constraint we have obtained on the field theory Lagrange space using positivity bounds is new and has no analog in the geometry of $\mathcal{M}$.

\subsection{Positivity Bounds at Higher Points}

From considering the four-derivative term, we now know that if the h-covariant derivative of the (v)hv-torsion is non-zero at the vacuum, it must be positive. What constraint is there were the four-derivative term to vanish? If so, the first non-zero higher-derivative interaction appears at some $n=2k$ (c.f. \cref{eqn:Lagy}):
\begin{equation}
\mathcal{L}(x, y) = V(x) + g_{\alpha \beta}(x)\, y^\alpha y^\beta + c_{\gamma_1 \ldots \gamma_{2k}} (x)\, y^{\gamma_1} \cdots y^{\gamma_{2k}} + \mathcal{O}(\partial^{2k+2}) \,.
\label{eqn:Lagy2k}
\end{equation}
Like in \cref{eqn:AffineCombination1}, we consider affine combinations of the Berwald and Cartan connections with h-Christoffel symbols:
\begin{equation}
F{^\alpha}_{\beta\gamma} = \left(1-b\right) \frac{\partial N{^\alpha}_\gamma}{\partial y^\beta} + b\, \frac12\, \mathfrak{g}^{\alpha\delta} \left( \frac{\delta \mathfrak{g}_{\delta\gamma}}{\delta x^\beta} + \frac{\delta \mathfrak{g}_{\delta\beta}}{\delta x^\gamma} - \frac{\delta \mathfrak{g}_{\beta\gamma}}{\delta x^\delta} \right) \,,
\end{equation}
leading to the (v)hv-torsion:
\begin{equation}
P{^\alpha}_{\beta\gamma} = F{^\alpha}_{\gamma\beta} - \frac{\partial N{^\alpha}_\beta}{\partial y^\gamma} = -b\, \frac{\partial N{^\alpha}_\beta}{\partial y^\gamma} + b\, \frac12\, \mathfrak{g}^{\alpha\delta} \left( \frac{\delta \mathfrak{g}_{\delta\gamma}}{\delta x^\beta} + \frac{\delta \mathfrak{g}_{\delta\beta}}{\delta x^\gamma} - \frac{\delta \mathfrak{g}_{\beta\gamma}}{\delta x^\delta} \right) \, .
\end{equation}
For $n>4$, this is zero on the null section. Nevertheless, an analog to \cref{eqn:vhvP4pt} can be obtained by taking additional v-covariant derivatives --- we find the first nonzero component on the null section to be\footnote{The choice of $C{^\alpha}_{\beta\gamma}$ does not matter for this result, as we are only computing the first nonzero v-covariant derivative on the null section.}
\begin{equation}
P_{\alpha_1 \alpha_2 \alpha_3} |_{\alpha_4 \ldots \alpha_{n-1}}(x,0) = -b\, \frac{n!}{8}\, c{^\delta}_{\alpha_1 \ldots \alpha_{n-1}}(x) V_{,\delta}(x) \,,
\end{equation}
and hence obtain an analog to \cref{eqn:c4asTorsion}:
\begin{equation}
\bar{P}_{\alpha_1 \alpha_2 \alpha_3} |_{\alpha_4 \ldots \alpha_{n-1} / \alpha_n} = \left[ \frac{\partial}{\partial x^{\alpha_n}}\, P_{\alpha_1 \alpha_2 \alpha_3} |_{\alpha_4 \ldots \alpha_{n-1}}(x,0) \right] \bigg|_{x=\bar{x}} = b\, \frac{ n!}{4}\, m^2_{\alpha_n}\, \bar{c}_{\alpha_1 \ldots \alpha_n} \,.
\label{eqn:c2kasTorsion}
\end{equation}
We can now extend the argument in \cref{subsec:FourPointPositivity} to higher-point amplitudes $\Amp_{n}^\rho$, generalizing~\cite{Chandrasekaran:2018qmx} to multi-scalar theories to obtain positivity bounds on the derivative of the (v)hv-torsion above.

\subsubsection*{The Case of Even $k$}

Suppose $k$ is even, and consider $n$-point scattering of the superposed states
\begin{equation}
\ket{i} = \ket{i+k} = \sum_\alpha \sqrt{2\bar{\mathfrak{g}}_{\alpha\alpha}}\, \rho_i^\alpha \ket{\phi_\alpha}
\quad\text{for each}\quad
i \le k \,,
\end{equation}
with forward momenta
\begin{equation}
p_i = - p_{i+k} 
\quad\text{for each}\quad
i \le k \,,
\end{equation}
and the alternating assignment
\begin{subequations}
\begin{align}
p_i &= p_1   \quad\text{for odd}\quad   i \le k \,,\\[3pt]
p_i &= p_2   \quad\text{for even}\quad  i \le k \,.
\end{align}
\end{subequations}
The forward amplitude $\Amp_n^\rho$ is then a function of the center-of-mass energy squared:
\begin{equation}
s \coloneqq (p_1 + \ldots + p_k)^2 = \frac{k^2}{2} \left(p_1 \cdot p_2\right) + \frac{k^2}{2}\, m^2 \,.
\end{equation}

The generalization of the integral in \cref{eqn:I4def} is
\begin{equation}
I_n \coloneqq \frac{1}{2 \pi i} \oint_\gamma \frac{ds}{s^{k+1}} \Amp_n^\rho(s) \,,
\end{equation}
which picks out the coefficient of the $s^k$ term in $\Amp_n^\rho(s)$ at low energies. The only tree-level diagram that contributes to this term is the $n$-point contact diagram --- vertices below $n$ points each contribute at most one power in $s$, as they arise from horizontal derivatives of $\mathfrak{g}$ evaluated at the vacuum. Counting then reveals that for a propagator diagram to contribute to $s^k$, it must contain a 3-point vertex. But we can always go to normal coordinates on the null section like in \cref{sec:horgeom} to make the order-$s$ part of the three-point vertex vanish. Thus, near $s = 0$ in these coordinates:
\begin{equation}
\Amp_n^\rho = \sum_\sigma (p_{\sigma_1} \cdot p_{\sigma_2}) \ldots (p_{\sigma_{n-1}} \cdot p_{\sigma_n})\,
\bar{c}_{\alpha_1 \ldots \alpha_n}\,
\rho_1^{\alpha_1} \ldots \rho_k^{\alpha_k}\,
\rho_1^{\alpha_{k+1}} \ldots \rho_k^{\alpha_n}
+ \mathcal{O}(s^{k-1}) \,.
\end{equation}
Here $\sigma$ runs over all permutations of $\{1, 2, \ldots, n\}$ and the symmetry of the Wilson coefficient has been applied. By our choice of kinematics, each pair of contracted momenta must contain one odd and one even index to contribute to $s^k$, and the product of $k$ pairs always yields a positive sign. Hence
\begin{equation}
I_n = \frac{(2^k k!)^2}{k^n}\, \bar{c}_{\alpha_1 \ldots \alpha_n}\,
\rho_1^{\alpha_1} \ldots \rho_k^{\alpha_k}\,
\rho_1^{\alpha_{k+1}} \ldots \rho_k^{\alpha_n} \,.
\end{equation}

We can again deform the contour and, assuming the Froissart-Martin bound to discard the boundary integral, we get
\begin{equation}
I_n = \frac{1}{2 \pi i} \left ( \int_{-\infty}^{0} + \int_{0}^{\infty} \right ) \frac{ds}{s^{k+1}}\, \text{disc} \, \Amp_n^\rho(s) \,.
\end{equation}
Making use of crossing symmetry, Hermitian analyticity and unitarity:
\begin{equation}
\Amp_n^\rho(-s) = \Amp_n^\rho \left( s + k^2 m^2 \right) \,, \quad \Amp_n^\rho(s^*) = \left[ \Amp_n^\rho(s) \right]^* \,, \quad \text{Im}\, \Amp_n^\rho(s) \ge 0 \,,
\end{equation}
we deduce that the integral must be non-negative:
\begin{equation}
I_n \ge 0 \,,
\end{equation}
\begin{equation}
\bar{P}_{\alpha_1 \alpha_2 \alpha_3} |_{\alpha_4 \ldots \alpha_{n-1} / \alpha_n}\,
\rho_1^{\alpha_1} \ldots \rho_k^{\alpha_k}\,
\rho_1^{\alpha_{k+1}} \ldots \rho_k^{\alpha_n} \ge 0 \,.
\label{eqn:Pkevenbound}
\end{equation}
In other words, the h-covariant derivative of the $(n-4)$th v-covariant derivative of the (v)hv-torsion is a positive semi-definite $n$-quadratic form.

The constraint \cref{eqn:Pkevenbound} can actually be improved by extending the argument to inelastic scattering~\cite{Zhang:2020jyn, Arkani-Hamed:2021ajd, Freytsis:2022aho}. Specifically, we consider the $n$-point amplitude $\Amp_{i\to f}$ with the initial and final states
\begin{subequations}\label{eqn:ifstates}
\begin{align}
\ket{i} &= \ket{\phi_{\alpha_1} \cdots \phi_{\alpha_k}} \,, \\[5pt]
\ket{f} &= \ket{\phi_{\alpha_{k+1}} \cdots \phi_{\alpha_n}} \,,
\end{align}
\end{subequations}
under the same kinematics as before. The integral
\begin{equation}
I_{i\to f} = \frac{1}{2 \pi i} \oint_\gamma \frac{ds}{s^{k+1}}\, \Amp_{i\to f}(s) \,,
\end{equation}
is proportional to $\bar{c}_{\alpha_1 \ldots \alpha_n}$ with a positive sign. Again, deforming the contour does not change the value of the integral, but crossing symmetry now implies
\begin{equation}
\Amp_{i\to f} (s) = \Amp_{f\to i} (s)
\qquad\text{and}\qquad
\Amp_{i\to f} (-s) = \Amp_{i'\to f'} \left( s + k^2 m^2 \right) \,,
\end{equation}
where in the latter, the new states after crossing are
\begin{align}
\ket{i'} &= \ket{\phi_{\alpha_1} \phi_{\alpha_{k+2}} \phi_{\alpha_3} \phi_{\alpha_{k+4}} \cdots \phi_{\alpha_{k-1}} \phi_{\alpha_n}} \,, \\[5pt]
\ket{f'} &= \ket{\phi_{\alpha_{k+1}} \phi_{\alpha_2} \phi_{\alpha_{k+3}} \phi_{\alpha_4} \cdots \phi_{\alpha_{n-1}} \phi_{\alpha_k}} \,.
\end{align}
Invoking the Froissart-Martin bound, Hermitian analyticity, and the generalized optical theorem, we find
\begin{align}
I_{i\to f} &= \frac{1}{2 \pi} \int_{0}^{\infty} \frac{ds}{s^{k+1}}\, \text{Im}\, \Big[ \Amp_{i\to f}(s) + \Amp_{f\to i}(s) + \Amp_{i'\to f'}\left(s + k^2 m^2\right) + \Amp_{f'\to i'}\left(s + k^2 m^2\right) \Big]
\notag\\[8pt]
&= \frac{1}{4\pi} \sum_X \int_{0}^{\infty} \frac{ds}{s^{k+1}} \Big[ \Amp_{i\to X}(s) \Amp_{f\to X}(s)^*
\notag\\[-3pt]
&\hspace{120pt}
+ \Amp_{i'\to X}\left(s + k^2 m^2\right) \Amp_{f'\to X}\left(s + k^2 m^2\right)^* + \text{c.c.} \Big] \,,
\end{align}
where $X$ runs over intermediate states. Decomposing the various $\mathcal{A}_{j \rightarrow X}$ into real and imaginary parts, the symmetry of $\bar{c}_{\alpha_1 \ldots \alpha_n}$ means that up to a mass factor, $\bar{P}_{\alpha_1 \alpha_2 \alpha_3} |_{\alpha_4 \ldots \alpha_{n-1} / \alpha_n}$ is a (generally continuous) sum of symmetrized outer products of rank-$k$ real tensors $t$, a stronger positivity bound:
\begin{equation}
\bar{P}_{\alpha_1 \alpha_2 \alpha_3} |_{\alpha_4 \ldots \alpha_{n-1} / \alpha_n} = m^2_{\alpha_n} \sum_t t_{(\alpha_1 \ldots \alpha_k} t_{\alpha_{k+1} \ldots \alpha_n)} \,.
\label{eq:higherptpostt}
\end{equation}

\subsubsection*{The Case of Odd $k$}

Now suppose $k$ is odd. For the same argument to hold, the contour integral must be defined to pick out an even power of $s$ in $\Amp_n^\rho(s)$. The largest such power is $k-1$, but even with this choice there will be contributions from trees comprising $k-1$ four-point vertices. Let us assume that $\bar{\mathcal{R}} = 0$ so that these diagrams vanish. The same setup for elastic scattering means that $p_k = -p_n = p_1$ and the highest power of $s$ in $\Amp_n^\rho(s)$ is $k-1$, arising solely from the $n$-point contact diagram. Here, the center-of-mass energy squared is
\begin{equation}
s = \frac{k^2-1}{2}\, (p_1 \cdot p_2) + \frac{k^2+1}{2}\, m^2 \,,
\end{equation}
and at low energies we have
\begin{equation}
I_{n, \text{ odd } k} = \frac{1}{2 \pi i} \oint_\gamma \frac{ds}{s^k}\, \Amp_n^\rho(s)
= k! \, (k+1)! \left ( \frac{4}{k^2-1} \right )^{k-1} m^2 \, \bar{c}_{\alpha_1 \ldots \alpha_n}\,
\rho_1^{\alpha_1} \ldots \rho_k^{\alpha_k}\,
\rho_1^{\alpha_{k+1}} \ldots \rho_k^{\alpha_n} \,.
\end{equation}
We again deform the integration contour $\gamma$ into $\gamma'$, and invoke the Froissart bound, crossing symmetry, Hermitian analyticity, and the optical theorem to obtain \cref{eqn:Pkevenbound}. The stronger result \cref{eq:higherptpostt} can also be obtained in a similar fashion. The difference in sign compared to~\cite{Chandrasekaran:2018qmx} is merely due to spacetime metric signature convention, and the generalization to $\bar{\mathcal{R}} \neq 0$ is straightforward: we add to $I_{n, \text{ odd } k}$ the contribution from trees with four-point vertices, and the positivity bound will apply to a sum of the Wilson coefficient and contractions of $\bar{\mathcal{R}}$.

\subsection{Unitarity Bounds as Size Constraints}

Another example of constraints on higher-derivative physics comes from unitarity bounds, which limit the size of the scattering amplitude and hence its leading-order term as momentum grows. On the other hand, this term is encoded in the (v)hv-torsion at the vacuum, so that the vertical geometry determines the energy scale up to which the effective field theory can be a unitary description for scattering.

Let us adapt the computational techniques in~\cite{Chang:2019vez, Abu-Ajamieh:2020yqi} and work out the unitarity bounds of a weakly-coupled Lagrangian whose loop corrections are negligible. Denoting the incoming and outgoing scattering states as $\ket{i}$ and $\bra{f}$, the unitarity of the $S$-matrix $S_{fi} = \delta_{fi} + i M_{fi}$ implies
\begin{equation}
\sum_{f\ne i} |M_{fi}|^2 \le 1
\quad\implies\quad
|M_{fi}| \le 1
\quad\text{for all}\quad
f \ne i \,.
\end{equation}
Specializing to four-point $s$-wave scattering of type
\begin{equation}
\ket{i} = \ket{\phi_{\alpha_1} \phi_{\alpha_2}}
\quad\rightarrow\quad
\ket{f} = \ket{\phi_{\alpha_3} \phi_{\alpha_4}} \,,
\label{eqn:4ptifstates}
\end{equation}
we adopt for computational purposes the center-of-mass frame and work with the center-of-mass energy $\sqrt{s}$ and the angle $\theta$ between three-momenta $\mathbf{p}_1$ and $\mathbf{p}_3$. Then we have
\begin{equation}
M_{fi} \propto \frac{1}{16\pi} \int_0^\pi d\theta \sin \theta\, \Amp_4(s, \theta) \,,
\label{eqn:Mswave}
\end{equation}
where the proportionality constant is $1$, $1 / \sqrt{2}$ or $1/2$ depending on whether the particles in each pair are distinguishable. Now, recall the covariant expression \cref{eqn:A4Original} for the four-point tree-level scattering amplitude --- the dominant piece when $s$ is large is that of order $s^2$:
\begin{equation}
\Amp_4(s, \theta) = 2 \left( s^2 + t^2 + u^2 \right) \frac{\bar{c}_{\alpha_1 \alpha_2 \alpha_3 \alpha_4}}{\prod_{i=1}^4 \sqrt{2\bar{\mathfrak{g}}_{\alpha_i\alpha_i}}} + \text{ subleading terms}\,,
\end{equation}
which holds regardless of the choice of connection in \cref{eqn:A4Original}. Taking the case that both pairs in \cref{eqn:4ptifstates} are indistinguishable particles (which corresponds to a proportionality constant $1/2$ in \cref{eqn:Mswave}), we get
\begin{equation}
M_{fi} = \frac{5}{24\pi}\, s^2\, \frac{\bar{c}_{\alpha_1 \alpha_2 \alpha_3 \alpha_4}}{\prod_{i=1}^4 \sqrt{2\bar{\mathfrak{g}}_{\alpha_i\alpha_i}}}
= \frac{5}{144\pi}\, \frac{s^2}{b\, m_{\alpha_4}^2}\, \frac{\bar{P}_{\alpha_1 \alpha_2 \alpha_3/\alpha_4}}{\prod_{i=1}^4 \sqrt{2\bar{\mathfrak{g}}_{\alpha_i\alpha_i}}} \,.
\end{equation}
So long as $f\ne i$, we find
\begin{equation}
|\bar{P}_{\alpha_1\alpha_2 \alpha_3 / \alpha_4}| \leq \frac{144 \pi}{5}\, \frac{|b|\, m_{\alpha_4}^2}{s^2}\, \prod_{i=1}^4 \sqrt{2\bar{\mathfrak{g}}_{\alpha_i \alpha_i}} \,.
\label{eqn:PbarUnitarity}
\end{equation}
Bearing in mind the symmetry of $\bar{P}_{\alpha_1\alpha_2 \alpha_3 / \alpha_4}$ in its first three indices, this is a unitarity bound on the size of every component apart from that with all indices equal. An immediate improvement follows from summing $|M_{fi}|^2$ over $f \neq i$. Combined with the four-point positivity bound, the unitarity bound tells us that the (v)hv-torsion cannot be increasing too fast at the vacuum --- the faster it increases, the smaller the maximum scale for which the effective field theory is valid.

Just like positivity bounds, unitarity also limits the size of higher-order derivatives of the (v)hv-torsion if the first non-zero Wilson coefficient appears at $n > 4$. Consider again the Lagrangian in \cref{eqn:Lagy2k}. In the $n$-point scattering amplitude of the type in \cref{eqn:ifstates}, the leading-order piece in momentum is given by
\begin{equation}
\Amp_n = \left ( -\frac{1}{2} \right )^k \sum_\sigma (s_{\sigma_1 \sigma_2}) \ldots (s_{\sigma_{n-1} \sigma_n}) \frac{\bar{c}_{\alpha_1 \ldots \alpha_n}}{\prod_{i=1}^n \sqrt{2\bar{\mathfrak{g}}_{\alpha_i \alpha_i}}} + \text{ subleading terms}\,,
\end{equation}
where $\sigma$ runs over all permutations of $\{1, 2, \cdots, n\}$. Taking the case that both the initial and final states comprise indistinguishable particles, the $s$-wave component is
\begin{equation}
M_{fi} = \frac{1}{k!}\, \frac{1}{\sqrt{\mathrm{Vol}_i \mathrm{Vol}_f}}\, \int \text{dLIPS}_i \int \text{dLIPS}_f \, \Amp_n \,.
\end{equation}
Here $\mathrm{Vol}_i$ ($\mathrm{Vol}_f$) denotes the Lorentz-invariant phase space volume of the state $i$ ($f$). For either $k$-body state, the phase space volume in the large-$s$ limit is
\begin{equation}
\mathrm{Vol}_k = \int \mathrm{dLIPS}_k = \frac{1}{8\pi\, (k-1)!\, (k-2)!} \left( \frac{s}{16\pi^2} \right)^{k-2} \,.
\end{equation}
Using this we get
\begin{align}
M_{fi} &= \frac{(k-1)!\, (k-2)!}{k!}\, (2\pi)^{n-3}\, (-2)^k\, J_n\, s^2\, \frac{n!}{4}\, \frac{\bar{c}_{\alpha_1 \ldots \alpha_n}}{\prod_{i=1}^n \sqrt{2\bar{\mathfrak{g}}_{\alpha_i\alpha_i}}}
\notag\\[8pt]
&= \frac{(k-2)!}{k}\, (2\pi)^{n-3}\, (-2)^k\, J_n\, s^2\, \frac{\bar{P}_{\alpha_1 \alpha_2 \alpha_3} |_{\alpha_4 \ldots \alpha_{n-1} / \alpha_n}}{b\, m_{\alpha_n}^2\, \prod_{i=1}^n \sqrt{2\bar{\mathfrak{g}}_{\alpha_i\alpha_i}}} \,,
\label{eqn:M2k}
\end{align}
where $J_n$ is a number defined by a purely kinematic integral:
\begin{equation}
J_n = \lim_{\text{large } s} \frac{1}{n!} \sum_\sigma \int \text{dLIPS}_i \int \text{dLIPS}_f \, \frac{s_{\sigma_1 \sigma_2}}{s} \ldots \frac{s_{\sigma_{n-1} \sigma_n}}{s} \,,
\end{equation}
with e.g. $J_4 = 5 / (576\pi^2)$. For $f\ne i$, \cref{eqn:M2k} then yields a unitarity bound:
\begin{equation}
| \bar{P}_{\alpha_1 \alpha_2 \alpha_3} |_{\alpha_4 \ldots \alpha_{n-1} / \alpha_n} | \le \frac{k}{(2\pi)^{n-3}\, 2^k\, (k-2)!\, |J_n|}\, \frac{|b|\, m^2_{\alpha_n}}{s^2}\, \prod_{i=1}^n \sqrt{2\bar{\mathfrak{g}}_{\alpha_i \alpha_i}} \,.
\end{equation}

At this point, we remark that the choice of coordinates is inconsequential to the key idea that positivity and unitarity bounds are synonymous with constraints on the vertical geometry of field theory Lagrange space. The normal diagonalized coordinates that we often work in are chosen to make the notion of positivity and magnitude precise yet simple.

By now, it should be clear that effective field theories that obey physical principles we cherish correspond to a class of Lagrange spaces with special vertical geometry at the vacuum --- a particular torsion component must be zero but in some sense increasing, and the rate of increase limits the scale of validity of the theory. The translation from physical to geometric constraints in the framework of Lagrange spaces is truly a new insight that has no counterpart in Riemannian geometry. By enlarging the manifold that we embed the Lagrangian into from $\mathcal{M}$ to $T\mathcal{M}$, there are now degrees of freedom to simultaneously encode more d-tensors' worth of information at one point. This is what makes Lagrange spaces a more powerful framework for understanding effective field theories.

\section{Conclusion and Outlook}
\label{sec:conc}

An effective field theory is an expansion in both fields and derivatives --- directions which a Lagrange space naturally distinguishes by assigning separate coordinates to fields, here denoted $x$, and their first derivatives, denoted $y$. We have demonstrated in this paper how Wilson coefficients in scalar field theories map to d-tensors on a Lagrange space. The d-tensors transform covariantly under non-derivative field redefinitions of the Lagrangian, and therefore combine by necessarily simple rules --- which we have enumerated --- to form the coefficients of field-redefinition-invariant scattering amplitudes.

Lagrange spaces have proven to be a powerful generalization of the existing Riemannian framework, which builds covariant tensors on the scalar field manifold charted only by $x$. While we ultimately restrict to $y=0$, the horizontal geometry (i.e., the physics in the ``$x$ directions'') of the Lagrange space at the vacuum benefits from a greater freedom in the choice of connection, corresponding to a freedom in the amplitude to shuffle physics between terms by momentum conservation. We found alternative connections to the Riemannian Levi-Civita connection that can better capture the physics at higher orders in momentum.

Meanwhile, the extra vertical degrees of freedom (the physics in the ``$y$ directions'') strictly accommodate Wilson coefficients of operators with four or more derivatives. Physical constraints on said operators stemming from positivity and perturbative unitarity are thus given a geometric interpretation in terms of components of the torsion of the manifold --- physical theories translate to a special class of vertical geometry.

Still, there is more to be understood about Lagrange spaces than what we have found so far. For example, the vertical geometry at the vacuum, which results from the full tower of higher-derivative operators, has yet to be fully characterized for a generic Lagrangian. We have highlighted the key phenomenon of a vanishing but increasing (v)hv-torsion resulting from positivity bounds, but limited our scope to the first non-vanishing Wilson coefficient. Positivity bounds involving multiple Wilson coefficients and their geometric description are promising topics for future investigation.

Additionally, there is a lot more information to be explored --- in principle the whole Lagrangian --- embedded in field theory Lagrange space beyond the vacuum. Geometric quantities take on complicated coordinate expressions at a generic point in Lagrange space, where $y\neq 0$; nevertheless, there may be physical insights to be gained. In particular, one could hope for a geometric description of scattering amplitudes that is truly canonical from the Lagrangian with no additional d-tensor structures. Notably, the fundamental tensor $\mathfrak{g}(x,y)$ contains all the information of the Lagrangian, apart from the potential term, and is strongly reminiscent of the momentum-dependent metric of \cite{Cheung:2022vnd}. In \cite{Cheung:2022vnd}, all field and derivative dependence is subsumed into a momentum-dependent metric, allowing the coefficients of amplitudes to be expressed as momentum-dependent curvatures. The technology of Lagrange spaces may shed light on underlying mathematical structures in amplitudes that make such expressions possible. Furthermore, in the same spirit as extensions to the basic Riemannian framework, one can also consider how the geometry of Lagrange spaces changes once loop effects are introduced, or how the framework can be extended to theories with higher spin.

Finally, there are motivations to look beyond the realm of Lagrange spaces. In particular, the present picture is only applicable to Lagrangians with flavor-symmetric Wilson coefficients, and does not exemplify the invariance of scattering amplitudes under derivative field redefinitions. Both of these limitations are due to the inability of a Lagrange space to accommodate the Lorentz structure of derivatives, as it cannot distinguish $\partial_\mu x$ for different $\mu$. One solution is to introduce more degrees of freedom to the tangent bundle, hinting at more general geometries such as jet bundles. Each extra handle can broaden our ability to delineate the large space of field transformations, and by extension, the intricate subset of field redefinition invariant configurations on which on-shell physics resides.

\acknowledgments

We would like to thank John Celestial, Grant Remmen, and Chia-Hsien Shen for useful discussions, and Tim Cohen for collaboration at early stages of this work. The work of NC and YL was supported in part by the U.S.\ Department of Energy under the grant DE-SC0011702 and performed in part at the Kavli Institute for Theoretical Physics, supported by the National Science Foundation under Grant No.~NSF PHY-1748958. The work of XL is supported by the U.S.~Department of Energy under grant number DE-SC0009919.

\appendix
\section*{Appendices}
\addcontentsline{toc}{section}{\protect\numberline{}Appendices}%

\section{The Curvature and Torsion of \textit{N}-linear Connections}
\label{app:torcur}

An $N$-linear connection $\nabla$, being an affine connection on $T\mathcal{M}$, has the usual invariants of torsion $T$ and curvature $R$. The curvature tensor is a map
\begin{equation}
R\, (X, Y)\, Z = \Big( [\nabla_X, \nabla_Y] - \nabla_{[X, Y]} \Big)\, Z \,,
\end{equation}
where $X$, $Y$, $Z$ and the output are tangent vectors on $T\mathcal{M}$. Working in the Berwald basis of horizontal $\delta / \delta x$ and vertical $\partial / \partial y$, we can specify whether each of the tangent vectors is horizontal or vertical. However, the fact that $\nabla$ respects the horizontal-vertical decomposition entails that there are actually only three independent d-tensor components $\mathcal{R}$, $\mathcal{P}$, and $\mathcal{Q}$~\cite{Miron:1994nvt}:
\begin{subequations}
\begin{alignat}{2}
R \left (\frac{\delta}{\delta x^\gamma}, \frac{\delta}{\delta x^\delta} \right ) \frac{\delta}{\delta x^\beta} &= \mathcal{R}\indices{^\alpha_{\beta\gamma\delta}}\, \frac{\delta}{\delta x^\alpha} \,,\qquad
R \left (\frac{\delta}{\delta x^\gamma}, \frac{\delta}{\delta x^\delta} \right ) \frac{\partial}{\partial y^\beta} &&= \mathcal{R}\indices{^\alpha_{\beta\gamma\delta}}\, \frac{\partial}{\partial y^\alpha}
\,, \\[5pt]
R \left (\frac{\delta}{\delta x^\gamma}, \frac{\partial}{\partial y^\delta} \right ) \frac{\delta}{\delta x^\beta} &= \mathcal{P}\indices{^\alpha_{\beta\gamma\delta}}\, \frac{\delta}{\delta x^\alpha} \,,\qquad
R \left (\frac{\delta}{\delta x^\gamma}, \frac{\partial}{\partial y^\delta} \right ) \frac{\partial}{\partial y^\beta} &&= \mathcal{P}\indices{^\alpha_{\beta\gamma\delta}}\, \frac{\partial}{\partial y^\alpha}
\,, \\[5pt]
R \left (\frac{\partial}{\partial y^\gamma}, \frac{\partial}{\partial y^\delta} \right ) \frac{\delta}{\delta x^\beta} &= \mathcal{Q}\indices{^\alpha_{\beta\gamma\delta}}\, \frac{\delta}{\delta x^\alpha} \,,\qquad
R \left (\frac{\partial}{\partial y^\gamma}, \frac{\partial}{\partial y^\delta} \right ) \frac{\partial}{\partial y^\beta} &&= \mathcal{Q}\indices{^\alpha_{\beta\gamma\delta}}\, \frac{\partial}{\partial y^\alpha}
\,.
\end{alignat}
\end{subequations}
The remaining possibilities follow from anti-symmetry in $X$ and $Y$. Each component reads:
\begin{subequations}
\begin{align}
\mathcal{R}\indices{^\alpha_{\beta\gamma\delta}} &=
\frac{\delta F\indices{^\alpha_\beta_\delta}}{\delta x^\gamma}
+ F\indices{^\alpha_\epsilon_\gamma} F\indices{^\epsilon_\beta_\delta}
+ C\indices{^\alpha_\beta_\epsilon} \frac{\delta N\indices{^\epsilon_\delta}}{\delta x^\gamma} 
- \left( \gamma \leftrightarrow \delta \right)
\,, \\[5pt]
\mathcal{P}\indices{^\alpha_{\beta\gamma\delta}} &= -\frac{\partial F^\alpha_{\;\beta\gamma}}{\partial y^\delta} + C\indices{^\alpha_{\beta \delta / \gamma}} + C\indices{^\alpha_\beta_\epsilon} \left ( F\indices{^\epsilon_\delta_\gamma} - \frac{\partial N\indices{^\epsilon_\gamma}}{\partial y^\delta} \right )
\,, \\[5pt]
\mathcal{Q}\indices{^\alpha_{\beta\gamma\delta}} &=
\frac{\partial C\indices{^\alpha_\beta_\delta}}{\partial y^\gamma}
+ C\indices{^\alpha_\epsilon_\gamma} C\indices{^\epsilon_\beta_\delta}
- \left( \gamma \leftrightarrow \delta \right)
\,.
\end{align}
\end{subequations}

Meanwhile, the torsion is given by
\begin{equation}
T\, (X, Y) = \nabla_X Y - \nabla_Y X - [X, Y] \,.
\end{equation}
Similarly, there are only five independent d-tensor components, one of which is the v-Christoffel symbol:
\begin{subequations}
\begin{align}
T \left (\frac{\delta}{\delta x^\beta}, \frac{\delta}{\delta x^\gamma} \right ) &= T\indices{^\alpha_{\beta\gamma}}\, \frac{\delta}{\delta x^\alpha}
+ R\indices{^\alpha_{\beta\gamma}}\, \frac{\partial}{\partial y^\alpha}
\,, \\[5pt]
T \left (\frac{\delta}{\delta x^\beta}, \frac{\partial}{\partial y^\gamma} \right ) &= -C\indices{^\alpha_{\beta\gamma}}\, \frac{\delta}{\delta x^\alpha}
+ P\indices{^\alpha_{\beta\gamma}}\, \frac{\partial}{\partial y^\alpha}
\,, \\[5pt]
T \left (\frac{\partial}{\partial y^\beta}, \frac{\partial}{\partial y^\gamma} \right ) &= S\indices{^\alpha_{\beta\gamma}}\, \frac{\partial}{\partial y^\alpha}
\,.
\end{align}
\end{subequations}
The components read:
\begin{subequations}\label{eqn:Torsion}
\begin{alignat}{2}
T\indices{^\alpha_{\beta\gamma}} &= F\indices{^\alpha_{\gamma\beta}} - \left( \beta \leftrightarrow \gamma \right)
\,,&\qquad
R\indices{^\alpha_{\beta\gamma}} &= \frac{\delta N\indices{^\alpha_\gamma}}{\delta x^\beta} - \left( \beta \leftrightarrow \gamma \right)
\,, \\[5pt]
P\indices{^\alpha_{\beta\gamma}} &= F\indices{^\alpha_{\gamma\beta}} - \frac{\partial N\indices{^\alpha_\beta}}{\partial y^\gamma}
\,,&\qquad
S\indices{^\alpha_{\beta\gamma}} &= C\indices{^\alpha_{\gamma\beta}} - \left( \beta \leftrightarrow \gamma \right)
\,.
\end{alignat}
\end{subequations}
where $R\indices{^\alpha_{\beta\gamma}}$ should not be confused with the hh-curvature.

\section{From Partial to Covariant Derivatives in Normal Coordinates}
\label{app:Vgccov}

Normal coordinates for a general affine connection $D$ at a point $\bar{x}$ on a manifold $\mathcal{M}$ can be constructed via the exponential map~\cite{Iliev:2006et}. For every tangent vector $y \in T_{\bar{x}} \mathcal{M}$, there is a unique geodesic $\gamma_y(t)$ under $D$ that originates from $\bar{x}$ at $t = 0$ with initial velocity $y$. The exponential map takes $y$ to $\gamma_y(1) \in \mathcal{M}$ if the latter exists. Restricting to appropriate neighborhoods of the origin in $T_{\bar{x}} \mathcal{M}$ and $\bar{x}$ in $\mathcal{M}$ makes this a diffeomorphism. We are then free to identify $T_{\bar{x}} \mathcal{M}$ with real coordinate space, thereby arriving at normal coordinates. For the scalar field manifold, the identification can be chosen to simultaneously diagonalize $g$ and the second derivative of $V$ at $\bar{x}$.

While the exponential map is not the only way to construct normal coordinates, it is the one we will use because partial derivatives of the Christoffel symbol $F$ then vanish under full symmetrization~\cite{Hatzinikitas:2000xe}:\footnote{Note that no property peculiar to the Levi-Civita connection was used in~\cite{Hatzinikitas:2000xe} to derive this result.}
\begin{equation}
\partial_{(\beta_1} \ldots \partial_{\beta_{k-1}} \bar{F}\indices{^\alpha_{\beta_k \beta_{k+1})}} \rightarrow 0 \quad\text{for all}\quad
k \geq 1 \,.
\end{equation}
Here, the symbol $\rightarrow$ is used to indicate an expression for the left-hand side that holds in normal coordinates. Composing covariant derivatives, we find:
\begingroup
\allowdisplaybreaks
\begin{subequations}\label{eq:VgcinD}
\begin{align}
\partial_{\beta_1} \ldots \partial_{\beta_k} \bar{V} &\rightarrow
D_{(\beta_1} \ldots D_{\beta_k)}\, \bar{V}
\,, \label{eq:VgcinD:V} \\[10pt]
\partial_{\beta_1} \ldots \partial_{\beta_k} \bar{g}_{\gamma_1 \gamma_2} &\rightarrow
D_{(\beta_1} \ldots D_{\beta_k)}\, \bar{g}_{\gamma_1 \gamma_2}
\notag\\[5pt]
&\quad
+ \partial_{(\beta_1} \ldots \partial_{\beta_{k-1}|} \left[ \bar{F}\indices{^\delta_{\gamma_1|\beta_k)}} \bar{g}_{\delta \gamma_2} + \bar{F}\indices{^\delta_{\gamma_2|\beta_k)}} \bar{g}_{\gamma_1 \delta} \right]
\notag\\[5pt]
&\quad
+ \partial_{(\beta_1} \ldots \partial_{\beta_{k-2}|} \left[ \bar{F}\indices{^\delta_{\gamma_1|\beta_{k-1}}} D_{\beta_k )} \bar{g}_{\delta \gamma_2} + \bar{F}\indices{^\delta_{\gamma_2|\beta_{k-1}}} D_{\beta_k )} \bar{g}_{\gamma_1 \delta} \right]
\notag\\[5pt]
&\quad
+ \cdots\cdots
\notag\\[5pt]
&\quad
+ \partial_{(\beta_1|} \left[ \bar{F}\indices{^\delta_{\gamma_1|\beta_2}} D_{\beta_3} \ldots D_{\beta_k )} \bar{g}_{\delta \gamma_2} + \bar{F}\indices{^\delta_{\gamma_2|\beta_2}} D_{\beta_3} \ldots D_{\beta_k )} \bar{g}_{\gamma_1 \delta} \right]
\,, \label{eq:VgcinD:g} \\[10pt]
\partial_{\beta_1} \ldots \partial_{\beta_k} \bar{c}_{\gamma_1 \gamma_2 \gamma_3 \gamma_4} &\rightarrow
D_{(\beta_1} \ldots D_{\beta_k)}\, \bar{c}_{\gamma_1 \gamma_2 \gamma_3 \gamma_4}
\notag\\[5pt]
&\quad
+ \partial_{(\beta_1} \ldots \partial_{\beta_{k-1}|} \left[ \bar{F}\indices{^\delta_{\gamma_1|\beta_k)}} \bar{c}_{\delta \gamma_2 \gamma_3 \gamma_4} + \bar{F}\indices{^\delta_{\gamma_2|\beta_k)}} \bar{c}_{\gamma_1 \delta \gamma_3 \gamma_4} + \ldots \right]
\notag\\[5pt]
&\quad
+ \partial_{(\beta_1} \ldots \partial_{\beta_{k-2}|} \left[ \bar{F}\indices{^\delta_{\gamma_1|\beta_{k-1}}} D_{\beta_k)} \bar{c}_{\delta \gamma_2 \gamma_3 \gamma_4} + \ldots \right]
\notag\\[5pt]
&\quad
+ \cdots\cdots
\notag\\[5pt]
&\quad
+ \partial_{(\beta_1|} \left[ \bar{F}\indices{^\delta_{\gamma_1|\beta_2}} D_{\beta_3} \ldots D_{\beta_k)} \bar{c}_{\delta \gamma_2 \gamma_3 \gamma_4} + \ldots \right]
\,. \label{eq:VgcinD:c}
\end{align}
\end{subequations}
\endgroup
Notice that the upper index $\delta$ of $F$ is never contracted with the index of $D$ as the result vanishes. Similarly, from the coordinate expression for the curvature tensor $\mathcal{R}$, we find
\begin{align}
\frac{k+1}{k-1}\, \partial_{(\beta_1} \ldots \partial_{\beta_{k-1}} \bar{F}\indices{^\alpha_{\beta_k) \gamma}} &\rightarrow
D_{(\beta_1} \ldots D_{\beta_{k-2}} \bar{\mathcal{R}}\indices{^\alpha_{\beta_{k-1} \beta_k) \gamma}}
\notag\\[5pt]
&\hspace{-80pt}
- \partial_{(\beta_1} \ldots \partial_{\beta_{k-2}} \left[ \bar{F}\indices{^\delta_{\beta_{k-1} | \gamma}} \bar{F}\indices{^\alpha_{\delta | \beta_k )}} \right]
\notag\\[5pt]
&\hspace{-80pt}
- \partial_{(\beta_1} \ldots \partial_{\beta_{k-3}|} \left[ \bar{F}\indices{^\alpha_{\delta|\beta_{k-2}}} \bar{\mathcal{R}}\indices{^\delta_{\beta_{k-1} \beta_k ) \gamma}}
- \left( \alpha \leftrightarrow \delta \text{ on } F \;,\; \delta \leftrightarrow \gamma \text{ on } \mathcal{R} \right) \right]
\notag\\[5pt]
&\hspace{-80pt}
- \partial_{(\beta_1} \ldots \partial_{\beta_{k-4}|} \left[ \bar{F}\indices{^\alpha_{\delta|\beta_{k-3}}} D_{\beta_{k-2} } \bar{\mathcal{R}}\indices{^\delta_{\beta_{k-1} \beta_k ) \gamma}}
- \left( \alpha \leftrightarrow \delta \text{ on } F \;,\; \delta \leftrightarrow \gamma \text{ on } \mathcal{R} \right) \right]
\notag\\[5pt]
&\hspace{-80pt}
- \cdots\cdots
\notag\\[5pt]
&\hspace{-80pt}
- \partial_{(\beta_1|} \left[ \bar{F}\indices{^\alpha_{\delta|\beta_2}} D_{\beta_3} \ldots D_{\beta_{k-2}} \bar{\mathcal{R}}\indices{^\delta_{\beta_{k-1} \beta_k ) \gamma}}
- \left(\alpha \leftrightarrow \delta \text{ on } F \;,\; \delta \leftrightarrow \gamma \text{ on } \mathcal{R} \right) \right]
\,.
\end{align}
If $F$ is symmetric, this equation can be used to recursively express partially symmetrized partial derivative of $F$ in terms of covariant derivatives of $\mathcal{R}$, which can then be substituted into \cref{eq:VgcinD:g,eq:VgcinD:c}. We can hence obtain exact covariant expressions for partial derivatives of $g$ and $c$ that are valid in normal coordinates. Suppressing any tensor products, \cref{eqn:VgcNormal} is recovered:
\begin{subequations}
\begin{align}
\bar{V}_{,\,\ldots} &\quad\longrightarrow\quad
\bar{V}_{/(\ldots)} \,, \\[5pt]
\bar{g}_{ij,\,\ldots} &\quad\longrightarrow\quad
\bar{g}_{ij/ (\ldots)} + \frac{r-3}{r-1}\, \Big[ \bar{\mathcal{R}}_{i(\ldots|j/|\ldots)} + \bar{\mathcal{R}}_{j(\ldots|i/|\ldots)} \Big]
+ \mathcal{O}(g_/ \mathcal{R},\, \mathcal{R}^2) \,, \\[5pt]
\bar{c}_{ijkl, \,\ldots} &\quad\longrightarrow\quad
\bar{c}_{ijkl / (\ldots)} + \mathcal{O}(c \mathcal{R}) \,.
\end{align}
\end{subequations}

\addcontentsline{toc}{section}{\protect\numberline{}References}%
\bibliographystyle{JHEP}
\bibliography{lagrange_JHEPRev}

\providecommand{\href}[2]{#2}\begingroup\raggedright\begin{thebibliography}{10}

\bibitem{Kamefuchi:1961sb}
S.~Kamefuchi, L.~O'Raifeartaigh and A.~Salam, \emph{{Change of variables and
  equivalence theorems in quantum field theories}},
  \href{https://doi.org/10.1016/0029-5582(61)90056-6}{\emph{Nucl. Phys.}
  {\bfseries 28} (1961) 529}.

\bibitem{Chisholm:1961tha}
J.S.R.~Chisholm, \emph{{Change of variables in quantum field theories}},
  \href{https://doi.org/10.1016/0029-5582(61)90106-7}{\emph{Nucl. Phys.}
  {\bfseries 26} (1961) 469}.

\bibitem{Coleman:1969sm}
S.R.~Coleman, J.~Wess and B.~Zumino, \emph{{Structure of phenomenological
  Lagrangians. 1.}},
  \href{https://doi.org/10.1103/PhysRev.177.2239}{\emph{Phys. Rev.} {\bfseries
  177} (1969) 2239}.

\bibitem{Arzt:1993gz}
C.~Arzt, \emph{{Reduced effective Lagrangians}},
  \href{https://doi.org/10.1016/0370-2693(94)01419-D}{\emph{Phys. Lett. B}
  {\bfseries 342} (1995) 189}
  [\href{https://arxiv.org/abs/hep-ph/9304230}{{\ttfamily hep-ph/9304230}}].

\bibitem{Meetz:1969as}
K.~Meetz, \emph{{Realization of chiral symmetry in a curved isospin space}},
  \href{https://doi.org/10.1063/1.1664881}{\emph{J. Math. Phys.} {\bfseries 10}
  (1969) 589}.

\bibitem{Honerkamp:1971sh}
J.~Honerkamp, \emph{{Chiral multiloops}},
  \href{https://doi.org/10.1016/0550-3213(72)90299-4}{\emph{Nucl. Phys. B}
  {\bfseries 36} (1972) 130}.

\bibitem{Honerkamp:1971xtx}
J.~Honerkamp and K.~Meetz, \emph{{Chiral-invariant perturbation theory}},
  \href{https://doi.org/10.1103/PhysRevD.3.1996}{\emph{Phys. Rev. D} {\bfseries
  3} (1971) 1996}.

\bibitem{Ecker:1971xko}
G.~Ecker and J.~Honerkamp, \emph{{Application of invariant renormalization to
  the nonlinear chiral invariant pion lagrangian in the one-loop
  approximation}},
  \href{https://doi.org/10.1016/0550-3213(71)90468-8}{\emph{Nucl. Phys. B}
  {\bfseries 35} (1971) 481}.

\bibitem{Alvarez-Gaume:1981exa}
L.~Alvarez-Gaume, D.Z.~Freedman and S.~Mukhi, \emph{{The Background Field
  Method and the Ultraviolet Structure of the Supersymmetric Nonlinear Sigma
  Model}}, \href{https://doi.org/10.1016/0003-4916(81)90006-3}{\emph{Annals
  Phys.} {\bfseries 134} (1981) 85}.

\bibitem{Alvarez-Gaume:1981exv}
L.~Alvarez-Gaume and D.Z.~Freedman, \emph{{Geometrical Structure and
  Ultraviolet Finiteness in the Supersymmetric Sigma Model}},
  \href{https://doi.org/10.1007/BF01208280}{\emph{Commun. Math. Phys.}
  {\bfseries 80} (1981) 443}.

\bibitem{Boulware:1981ns}
D.G.~Boulware and L.S.~Brown, \emph{{SYMMETRIC SPACE SCALAR FIELD THEORY}},
  \href{https://doi.org/10.1016/0003-4916(82)90192-0}{\emph{Annals Phys.}
  {\bfseries 138} (1982) 392}.

\bibitem{Howe:1986vm}
P.S.~Howe, G.~Papadopoulos and K.S.~Stelle, \emph{{The Background Field Method
  and the Nonlinear $\sigma$ Model}},
  \href{https://doi.org/10.1016/0550-3213(88)90379-3}{\emph{Nucl. Phys. B}
  {\bfseries 296} (1988) 26}.

\bibitem{Dixon:1989fj}
L.J.~Dixon, V.~Kaplunovsky and J.~Louis, \emph{{On Effective Field Theories
  Describing (2,2) Vacua of the Heterotic String}},
  \href{https://doi.org/10.1016/0550-3213(90)90057-K}{\emph{Nucl. Phys. B}
  {\bfseries 329} (1990) 27}.

\bibitem{Alonso:2015fsp}
R.~Alonso, E.E.~Jenkins and A.V.~Manohar, \emph{{A Geometric Formulation of
  Higgs Effective Field Theory: Measuring the Curvature of Scalar Field
  Space}}, \href{https://doi.org/10.1016/j.physletb.2016.01.041}{\emph{Phys.
  Lett. B} {\bfseries 754} (2016) 335}
  [\href{https://arxiv.org/abs/1511.00724}{{\ttfamily 1511.00724}}].

\bibitem{Cohen:2020xca}
T.~Cohen, N.~Craig, X.~Lu and D.~Sutherland, \emph{{Is SMEFT Enough?}},
  \href{https://doi.org/10.1007/JHEP03(2021)237}{\emph{JHEP} {\bfseries 03}
  (2021) 237} [\href{https://arxiv.org/abs/2008.08597}{{\ttfamily
  2008.08597}}].

\bibitem{Alonso:2016oah}
R.~Alonso, E.E.~Jenkins and A.V.~Manohar, \emph{{Geometry of the Scalar
  Sector}}, \href{https://doi.org/10.1007/JHEP08(2016)101}{\emph{JHEP}
  {\bfseries 08} (2016) 101}
  [\href{https://arxiv.org/abs/1605.03602}{{\ttfamily 1605.03602}}].

\bibitem{Nagai:2019tgi}
R.~Nagai, M.~Tanabashi, K.~Tsumura and Y.~Uchida, \emph{{Symmetry and geometry
  in a generalized Higgs effective field theory: Finiteness of oblique
  corrections versus perturbative unitarity}},
  \href{https://doi.org/10.1103/PhysRevD.100.075020}{\emph{Phys. Rev. D}
  {\bfseries 100} (2019) 075020}
  [\href{https://arxiv.org/abs/1904.07618}{{\ttfamily 1904.07618}}].

\bibitem{Helset:2020yio}
A.~Helset, A.~Martin and M.~Trott, \emph{{The Geometric Standard Model
  Effective Field Theory}},
  \href{https://doi.org/10.1007/JHEP03(2020)163}{\emph{JHEP} {\bfseries 03}
  (2020) 163} [\href{https://arxiv.org/abs/2001.01453}{{\ttfamily
  2001.01453}}].

\bibitem{Cohen:2021ucp}
T.~Cohen, N.~Craig, X.~Lu and D.~Sutherland, \emph{{Unitarity violation and the
  geometry of Higgs EFTs}},
  \href{https://doi.org/10.1007/JHEP12(2021)003}{\emph{JHEP} {\bfseries 12}
  (2021) 003} [\href{https://arxiv.org/abs/2108.03240}{{\ttfamily
  2108.03240}}].

\bibitem{Alonso:2021rac}
R.~Alonso and M.~West, \emph{{Roads to the Standard Model}},
  \href{https://doi.org/10.1103/PhysRevD.105.096028}{\emph{Phys. Rev. D}
  {\bfseries 105} (2022) 096028}
  [\href{https://arxiv.org/abs/2109.13290}{{\ttfamily 2109.13290}}].

\bibitem{Alonso:2022ffe}
R.~Alonso and M.~West, \emph{{On the effective action for scalars in a general
  manifold to any loop order}},
  \href{https://arxiv.org/abs/2207.02050}{{\ttfamily 2207.02050}}.

\bibitem{Helset:2022tlf}
A.~Helset, E.E.~Jenkins and A.V.~Manohar, \emph{{Geometry in scattering
  amplitudes}}, \href{https://doi.org/10.1103/PhysRevD.106.116018}{\emph{Phys.
  Rev. D} {\bfseries 106} (2022) 116018}
  [\href{https://arxiv.org/abs/2210.08000}{{\ttfamily 2210.08000}}].

\bibitem{Helset:2022pde}
A.~Helset, E.E.~Jenkins and A.V.~Manohar, \emph{{Renormalization of the
  Standard Model Effective Field Theory from geometry}},
  \href{https://doi.org/10.1007/JHEP02(2023)063}{\emph{JHEP} {\bfseries 02}
  (2023) 063} [\href{https://arxiv.org/abs/2212.03253}{{\ttfamily
  2212.03253}}].

\bibitem{Finn:2019aip}
K.~Finn, S.~Karamitsos and A.~Pilaftsis, \emph{{Frame Covariance in Quantum
  Gravity}}, \href{https://doi.org/10.1103/PhysRevD.102.045014}{\emph{Phys.
  Rev. D} {\bfseries 102} (2020) 045014}
  [\href{https://arxiv.org/abs/1910.06661}{{\ttfamily 1910.06661}}].

\bibitem{Finn:2020nvn}
K.~Finn, S.~Karamitsos and A.~Pilaftsis, \emph{{Frame covariant formalism for
  fermionic theories}},
  \href{https://doi.org/10.1140/epjc/s10052-021-09360-w}{\emph{Eur. Phys. J. C}
  {\bfseries 81} (2021) 572}
  [\href{https://arxiv.org/abs/2006.05831}{{\ttfamily 2006.05831}}].

\bibitem{Cohen:2022uuw}
T.~Cohen, N.~Craig, X.~Lu and D.~Sutherland, \emph{{On-Shell Covariance of
  Quantum Field Theory Amplitudes}},
  \href{https://doi.org/10.1103/PhysRevLett.130.041603}{\emph{Phys. Rev. Lett.}
  {\bfseries 130} (2023) 041603}
  [\href{https://arxiv.org/abs/2202.06965}{{\ttfamily 2202.06965}}].

\bibitem{Cheung:2022vnd}
C.~Cheung, A.~Helset and J.~Parra-Martinez, \emph{{Geometry-kinematics
  duality}}, \href{https://doi.org/10.1103/PhysRevD.106.045016}{\emph{Phys.
  Rev. D} {\bfseries 106} (2022) 045016}
  [\href{https://arxiv.org/abs/2202.06972}{{\ttfamily 2202.06972}}].

\bibitem{Kern1974}
J.~Kern, \emph{Lagrange geometry},
  \href{https://doi.org/10.1007/BF01238702}{\emph{Archiv der Mathematik}
  {\bfseries 25} (1974) 438}.

\bibitem{Miron:1994nvt}
R.~Miron and M.~Anastasiei, \emph{{The Geometry of Lagrange Spaces: Theory and
  Applications}}, Fundam.Theor.Phys., Springer Netherlands, Dordrecht (1994),
  \href{https://doi.org/10.1007/978-94-011-0788-4}{10.1007/978-94-011-0788-4}.

\bibitem{Antonelli:1996dq}
P.L.~Antonelli and R.~Miron, eds., \emph{{Lagrange and Finsler geometry:
  Applications to physics and biology}}, Kluwer, Dordrecht (1996),
  \href{https://doi.org/10.1007/978-94-015-8650-4}{10.1007/978-94-015-8650-4}.

\bibitem{Pham:1985cr}
T.N.~Pham and T.N.~Truong, \emph{{Evaluation of the Derivative Quartic Terms of
  the Meson Chiral Lagrangian From Forward Dispersion Relation}},
  \href{https://doi.org/10.1103/PhysRevD.31.3027}{\emph{Phys. Rev. D}
  {\bfseries 31} (1985) 3027}.

\bibitem{Ananthanarayan:1994hf}
B.~Ananthanarayan, D.~Toublan and G.~Wanders, \emph{{Consistency of the chiral
  pion pion scattering amplitudes with axiomatic constraints}},
  \href{https://doi.org/10.1103/PhysRevD.51.1093}{\emph{Phys. Rev. D}
  {\bfseries 51} (1995) 1093}
  [\href{https://arxiv.org/abs/hep-ph/9410302}{{\ttfamily hep-ph/9410302}}].

\bibitem{Adams:2006sv}
A.~Adams, N.~Arkani-Hamed, S.~Dubovsky, A.~Nicolis and R.~Rattazzi,
  \emph{{Causality, analyticity and an IR obstruction to UV completion}},
  \href{https://doi.org/10.1088/1126-6708/2006/10/014}{\emph{JHEP} {\bfseries
  10} (2006) 014} [\href{https://arxiv.org/abs/hep-th/0602178}{{\ttfamily
  hep-th/0602178}}].

\bibitem{Cornwall:1974km}
J.M.~Cornwall, D.N.~Levin and G.~Tiktopoulos, \emph{{Derivation of Gauge
  Invariance from High-Energy Unitarity Bounds on the s Matrix}},
  \href{https://doi.org/10.1103/PhysRevD.10.1145}{\emph{Phys. Rev. D}
  {\bfseries 10} (1974) 1145}.

\bibitem{Lee:1977eg}
B.W.~Lee, C.~Quigg and H.B.~Thacker, \emph{{Weak Interactions at Very
  High-Energies: The Role of the Higgs Boson Mass}},
  \href{https://doi.org/10.1103/PhysRevD.16.1519}{\emph{Phys. Rev. D}
  {\bfseries 16} (1977) 1519}.

\bibitem{Lee:1977yc}
B.W.~Lee, C.~Quigg and H.B.~Thacker, \emph{{The Strength of Weak Interactions
  at Very High-Energies and the Higgs Boson Mass}},
  \href{https://doi.org/10.1103/PhysRevLett.38.883}{\emph{Phys. Rev. Lett.}
  {\bfseries 38} (1977) 883}.

\bibitem{Dicus:1973gbw}
D.A.~Dicus and V.S.~Mathur, \emph{{Upper bounds on the values of masses in
  unified gauge theories}},
  \href{https://doi.org/10.1103/PhysRevD.7.3111}{\emph{Phys. Rev. D} {\bfseries
  7} (1973) 3111}.

\bibitem{Chanowitz:1985hj}
M.S.~Chanowitz and M.K.~Gaillard, \emph{{The TeV Physics of Strongly
  Interacting W's and Z's}},
  \href{https://doi.org/10.1016/0550-3213(85)90580-2}{\emph{Nucl. Phys. B}
  {\bfseries 261} (1985) 379}.

\bibitem{Finn:2018cfs}
K.~Finn, S.~Karamitsos and A.~Pilaftsis, \emph{{Eisenhart lift for field
  theories}}, \href{https://doi.org/10.1103/PhysRevD.98.016015}{\emph{Phys.
  Rev. D} {\bfseries 98} (2018) 016015}
  [\href{https://arxiv.org/abs/1806.02431}{{\ttfamily 1806.02431}}].

\bibitem{Iliev:2006et}
B.Z.~Iliev, \emph{{Handbook of normal frames and coordinates}}, Birkhäuser
  Basel (10, 2006), [\href{https://arxiv.org/abs/math/0610037}{{\ttfamily
  math/0610037}}].

\bibitem{Chaichian:1987zt}
M.~Chaichian and J.~Fischer, \emph{{Higher Dimensional Space-time and Unitarity
  Bound on the Scattering Amplitude}},
  \href{https://doi.org/10.1016/0550-3213(88)90394-X}{\emph{Nucl. Phys. B}
  {\bfseries 303} (1988) 557}.

\bibitem{Miramontes:1999gd}
J.L.~Miramontes, \emph{{Hermitian analyticity versus real analyticity in
  two-dimensional factorized S matrix theories}},
  \href{https://doi.org/10.1016/S0370-2693(99)00390-1}{\emph{Phys. Lett. B}
  {\bfseries 455} (1999) 231}
  [\href{https://arxiv.org/abs/hep-th/9901145}{{\ttfamily hep-th/9901145}}].

\bibitem{Remmen:2019cyz}
G.N.~Remmen and N.L.~Rodd, \emph{{Consistency of the Standard Model Effective
  Field Theory}}, \href{https://doi.org/10.1007/JHEP12(2019)032}{\emph{JHEP}
  {\bfseries 12} (2019) 032}
  [\href{https://arxiv.org/abs/1908.09845}{{\ttfamily 1908.09845}}].

\bibitem{Chandrasekaran:2018qmx}
V.~Chandrasekaran, G.N.~Remmen and A.~Shahbazi-Moghaddam, \emph{{Higher-Point
  Positivity}}, \href{https://doi.org/10.1007/JHEP11(2018)015}{\emph{JHEP}
  {\bfseries 11} (2018) 015}
  [\href{https://arxiv.org/abs/1804.03153}{{\ttfamily 1804.03153}}].

\bibitem{Zhang:2020jyn}
C.~Zhang and S.-Y.~Zhou, \emph{{Convex Geometry Perspective on the (Standard
  Model) Effective Field Theory Space}},
  \href{https://doi.org/10.1103/PhysRevLett.125.201601}{\emph{Phys. Rev. Lett.}
  {\bfseries 125} (2020) 201601}
  [\href{https://arxiv.org/abs/2005.03047}{{\ttfamily 2005.03047}}].

\bibitem{Arkani-Hamed:2021ajd}
N.~Arkani-Hamed, Y.-t.~Huang, J.-Y.~Liu and G.N.~Remmen, \emph{{Causality,
  unitarity, and the weak gravity conjecture}},
  \href{https://doi.org/10.1007/JHEP03(2022)083}{\emph{JHEP} {\bfseries 03}
  (2022) 083} [\href{https://arxiv.org/abs/2109.13937}{{\ttfamily
  2109.13937}}].

\bibitem{Freytsis:2022aho}
M.~Freytsis, S.~Kumar, G.N.~Remmen and N.L.~Rodd, \emph{{Multifield Positivity
  Bounds for Inflation}},  \href{https://arxiv.org/abs/2210.10791}{{\ttfamily
  2210.10791}}.

\bibitem{Chang:2019vez}
S.~Chang and M.A.~Luty, \emph{{The Higgs Trilinear Coupling and the Scale of
  New Physics}}, \href{https://doi.org/10.1007/JHEP03(2020)140}{\emph{JHEP}
  {\bfseries 03} (2020) 140}
  [\href{https://arxiv.org/abs/1902.05556}{{\ttfamily 1902.05556}}].

\bibitem{Abu-Ajamieh:2020yqi}
F.~Abu-Ajamieh, S.~Chang, M.~Chen and M.A.~Luty, \emph{{Higgs coupling
  measurements and the scale of new physics}},
  \href{https://doi.org/10.1007/JHEP07(2021)056}{\emph{JHEP} {\bfseries 07}
  (2021) 056} [\href{https://arxiv.org/abs/2009.11293}{{\ttfamily
  2009.11293}}].

\bibitem{Hatzinikitas:2000xe}
A.~Hatzinikitas, \emph{{A Note on Riemann normal coordinates}},
  \href{https://arxiv.org/abs/hep-th/0001078}{{\ttfamily hep-th/0001078}}.

\end{thebibliography}\endgroup

\end{document}